\documentclass[useAMS,usenatbib]{mn2e}
\usepackage{psfig}
\usepackage{rotating}
\usepackage{amsmath}
\title[SuperWASP-N Extra-Solar Planet candidates 3hr $<$ RA $<$ 6hr]{SuperWASP-North Extra-solar Planet Candidates between 3hr $<$ RA $<$ 6hr}

\author[W.I. Clarkson et al.]{W.I. Clarkson$^{1,2},$
B. Enoch$^{1}$,
C. A. Haswell$^{1}$,
A. J. Norton$^{1}$,
D.J. Christian$^{3}$,
\newauthor
A. Collier Cameron$^{4}$,
S.R. Kane$^{4,5}$,
K.D. Horne$^{4}$,
T.A. Lister$^{4,6,7}$,
R.A. Street$^{3,8}$,
\newauthor
R. G. West$^{9}$,
D. M. Wilson$^{6}$
N. Evans$^{6}$,
A. Fitzsimmons$^{3}$,
C. Hellier$^{6}$,
S.T. Hodgkin$^{10}$, 
\newauthor
J. Irwin$^{10}$,
F.P. Keenan$^{3}$,
J. P. Osborne$^{9}$,
N.R. Parley$^{1}$,
D.L. Pollacco$^{3}$,
R. Ryans$^{3}$,
\newauthor
I. Skillen$^{11}$,
and P.J. Wheatley$^{12}$ \\
$^{1}$Department of Physics \& Astronomy, The Open University, Milton Keynes, MK7 6AA, UK  \\
$^{2}$Space Telescope Science Institute, 3700 San Martin Drive, Baltimore, 21218, USA \\
$^{3}$APS Division, Department of Physics and Astronomy, Queen's University Belfast, Belfast, BT7 1NN, UK\\ 
$^{4}$SUPA, School of Physics \& Astronomy, University of St. Andrews, North Haugh, St. Andrews, Fife, KY16 9SS, UK  \\
$^{5}$University of Florida, PO~Box~112005, 211 Bryant Space Science
Center, Gainesville, FL, USA.\\
$^{6}$Astrophysics Group, School of Chemistry \& Physics, Keele University, Staffordshire, ST5 5BG, UK  \\
$^{7}$Las Cumbres Observatory, 6740B Cortona Drive, CA 93117, USA.\\
$^{8}$Dept. of Physics, Broida Hall, University of California, Santa
Barbara, CA 93106-9530, USA.\\
$^{9}$Department of Physics \& Astronomy, University of Leicester, Leicester, LE1 7RH, UK \\
$^{10}$Institute of Astronomy, University of Cambridge, Madingley Road, Cambridge, CB3 0HA, UK \\ 
$^{11}$Isaac Newton Group of Telescopes, Apartado de correos 321,
E-3700 Santa Cruz de la Palma, Tenerife, Spain\\ 
$^{12}$Department of Physics, University of Warwick, Coventry, CV4 7AL, UK \\
}
\date{Accepted 2007 July 27. Received 2007 July 27; in original form 2007 March 2}

\pagerange{\pageref{firstpage}--\pageref{lastpage}} \pubyear{2006}

\begin{document}

\maketitle

\label{firstpage}

\begin{abstract}
The Wide Angle Search for Planets (WASP) photometrically surveys a
large number of nearby stars to uncover candidate extrasolar planet
systems by virtue of small-amplitude lightcurve dips on a $\la 5$-day
timescale typical of the ``Hot-Jupiters.'' Observations with the
SuperWASP-North instrument between April and September 2004 produced a
rich photometric dataset of some 1.3$\times 10^9$ datapoints from 6.7
million stars. Our custom-built data acquisition and processing system
produces $\sim 0.02$ mag photometric precision at $V=$13.

We present the transit-candidates in the 03h-06h RA range. Of 141,895
lightcurves with sufficient sampling to provide adequate coverage,
2688 show statistically significant transit-like periodicities. Of
these, 44 pass visual inspection of the lightcurve, of which 24 are
removed through a set of cuts on the statistical significance of
artefacts. All but 4 of the remaining 20 objects are removed when
prior information at higher spatial-resolution from existing
catalogues is taken into account. Of the four candidates remaining,
one is considered a good candidate for follow-up observations with
three further second-priority targets. We provide detailed information
on these candidates, as well as a selection of the false-positives and
astrophysical false-alarms that were eliminated, and discuss briefly
the impact of sampling on our results.
\end{abstract}


%\correspondence{clarkson@stsci.edu}

\section{Introduction}

The discovery of radial-velocity variations indicative of a close
planetary companion to 51 Peg (Mayor \& Queloz 1995) caused a
revolution in studies of planetary formation and evolution, as planets
were traditionally thought not to exist as close as 0.05 A.U. to the
parent star (Pollack 1996). Subsequent radial-velocity searches have
uncovered 248 extrasolar planets (as of this
writing)\footnote{http://exoplanets.eu/catalog.php} orbiting
main-sequence objects (e.g. Udry et al. 2000).
%The
%ever-increasing timebase of the radial-velocity searches is producing
%detected systems with ever-larger periods (e.g. 55 Cnc d; Marcy et
%al. 2002). 
Many of these systems comprise a population with periods typically $<$
4d and orbital separations of order 0.05 A.U., and this was an early
challenge to theories of planet-formation and evolution.  

Transits combined with radial-velocity measurements offer the only
method to probe the internal structure of the exoplanets as they allow
the planetary radius and mass to be determined. 22 transiting
extrasolar planets\footnote[1]{including the two transiting exoplanets
  WASP-1b \& WASP-2b, which were discovered in other fields from the
  2004 WASP survey (Collier Cameron et al. 2007).} now have reported
mass \& radius estimates (Charbonneau et al. 2007b; Bakos et al. 2007b
submitted; Burke et al. 2007 submitted; Torres et
al. 2007\footnote[2]{as of this writing, this reference is not yet
  available on the astro-ph preprint server but can be found at
  http://exoplanets.eu/catalog-transit.php}), and although the number
of systems is still low, the emerging picture is of a ``main
sequence'' of gas-giants along the $\bar{\rho} \sim1.0$g cm$^3$ line
at masses $\ga$ 1 M$_J$, and a second, more diverse population at
lower mass but possibly inflated radius (e.g. Bakos et al 2007a).
%The currently-known population
%of transiting exoplanets can also be divided by planet-mass, one
%population at $\la 0.7M_J$ and the other $\ga 1.1 M_J$, with only
%XO-1b bridging the gap (McCullough et al. 2006). 

Studies of transiting exoplanets are driving current planetary
formation and disk-migration theory. Chi-squared fitting of
physically-motivated lightcurve models to the transit lightcurve
allows joint constraints on the orbital inclination and planetary
radius as a fraction of the stellar radius. The inclination estimate
from the transit-fitting then allows the planetary mass to be
estimated directly from radial-velocity measurements (e.g. Moutou et
al. 2006). The accuracy to which the planetary radius itself can be
determined is limited both by the photometric precision of the
lightcurve and the precision of the stellar-radius determination. The
latter is typically the limiting factor for space-based photometry
(Brown et al. 2001).

%section 2 
\section{Instrumention and Observations} %\label{instrobs} 
\subsection{Instrumentation}
SuperWASP-North (hereafter SW-N) was the first multi-camera WASP
instrument to enter operation. Full details can be found in Pollacco
et al (2006); we summarise here the features relevant to this
work. During 2004, the SW-N facility consisted of five wide-angle
(7.8$\degr$ $\times$ 7.8$\degr$ field of view) cameras on a rapid-slew
fork-mount that allows overheads (for slew and settling between
exposures) to be as short as 30 seconds even for slews $\ga
8\degr$. The 2048$\times$2048-pixel detectors yield a plate-scale
$\simeq 13.7''$/pixel, requiring careful consideration of the field
location and observation-depth to avoid washout by crowding. In 2004
the detector was unfiltered to maximise throughput, with an
instrumental bandpass covering most of the Johnson $VRI$ range, with
blue and red cutoffs at 4000 \AA \ and 10,000 \AA
\ respectively. While the mount pointing error is at most two pixels
rms across the sky, a slight misalignment of the instrument polar axis
leads to a position-drift of $\sim$ 10 pixels during the night.

\subsection{Observational Strategy} 

The WASP survey was planned around a broad-but-shallow approach to
maximise planet yield, as this brings three key benefits when
searching for exoplanet transits.

{\bf (i) - Further Exoplanet Diagnostics:} Detectable planetary
transits offer the possibility of probing the atmosphere of the
transiting planet. Charbonneau et al. (2006) list seven further
constraints that can be made on a transiting planet-star system, but
{\it only} if the parent star is sufficiently bright to allow high
enough signal-to-noise, including the setting of upper limits on
atmospheric absorption features (Deming et al. 2005b), the setting of
constraints on the vertical extent of the atmosphere by atomic species
(Vidal-Madjar et al. 2004; Charbonneau et al. 2002), the search for
spectroscopic features from the planet itself during secondary eclipse
(Richardson et al. 2003) and direct detection of thermal emission from
the planet itself (Deming et al 2005a). For the scientific return of
transiting exoplanets to be fully realised, then, ground-based transit
surveys such as WASP are typically optimised for objects at $V \la
13$.

{\bf (ii) - Facility of Follow-Up Observations:} One of the byproducts
of the {\it OGLE} microlensing project was a set of objects showing
apparent characteristics of exoplanet occultation
($P_{orb}\sim1$-$10$d, flux removal $\Delta F / F \sim1\%$, event
duration $\sim$hours; Udalski et al. 2002a-c). Strenuous follow-up
spectroscopic observations by several groups (e.g. Bouchy et al. 2005)
showed that a high fraction of these objects were astrophysical
false-alarms such as grazing-incidence stellar binaries or a
large-amplitude variable blended with the brighter target. Dedicated
narrow-deep photometric surveys (with e.g. {\it HST} or the upcoming
{\it Kepler} mission) afford such high coverage and spatial resolution
that this class of astrophysical false-alarm can be minimised to high
confidence from the photometry alone (to the level where $<$ 1
astrophysical false-positive is expected from the entire survey,
e.g. Sahu et al. 2006). For ground-based surveys, however, the
astrophysical false-positives will, for the foreseeable future, be a
large and important class of candidates; a population study using the
2MASS catalogue suggests an {\it astrophysical} false alarm to transit
ratio of at least 10:1 (Brown 2003); ground-based follow-up
observations are thus still essential. At the time of survey planning,
consideration of a variety of ground-based photometric observing
strategies (in the presence of uncorrelated (''white'')
noise; Horne 2003) suggested the SW-N hardware would provide survey
statistics competitive with all other existing transit surveys while
avoiding excessive crowding at fainter magnitudes. The SW-N limiting
magnitude to transits of $V\sim13$ (with 30-second exposures) allows
follow-up observations to take place with $\sim1/10$ the exposure time
(or collecting area) as similar observations of {\it OGLE} candidates
(themselves in the range $15 \le V \le 21$; Udalski et al. 2002a).

{\bf (iii) - Catalogue-based elimination of Astrophysical
  False-positives:} With the availability of the USNO-B1.0 (hereafter
USNO), Tycho-2 and 2MASS catalogues, multicolour absolute
magnitude-estimates already exist at higher spatial resolution than
the program variability observations. Tycho-2 is $\sim90\%$ complete
down to $V\simeq$ 11.5 mag (H$\o$g et al. 2000), while comparison with
the Sloan Digital Sky Survey suggests USNO is 97\% complete for stars
out to $g'$ $\sim$20 (roughly Johnson B $\sim 20$; Monet et
al. 2003). This allows obvious astrophysical false-positives to be
eliminated during analysis of the photometry; for the fields we report
here roughly 77\% of photometrically-promising candidates are ruled
out in this manner before any follow-up observations take place.

Thirty-second snapshots of each field of view are taken in sequences
of eight surrounding the Meridian; once the sequence is complete the
camera returns to the start of the sequence for the next run. The
rapid-slew capability of the mount allows a cadence of $\simeq$ 9
minutes per field. Fields centred at declination
$+23\degr \le \delta\degr \le +32\degr$ were generally chosen to
provide optimal survey grasp without crowding washout, though with
Galactic-plane avoidance some fields at other declinations were
sampled (see Pollacco et al. 2006)\footnote[2]{The full range of
  declinations imaged including galactic-plane avoidance is thus
  ($+12\degr < \delta\degr < +47\degr$)}. As the westernmost field in
the group of eight moves to high airmass, this field is abandoned and
a new field added on the east. The net result is a lightcurve with
$\sim9$-minute cadence, consisting of roughly 35 frames per night for
well-sampled fields. As the sky precesses throughout the year, roughly
60 nights' data are collected for each field for each camera.

\section{Analysis Techniques} 

We outline briefly the analysis techniques used in this project. The
reduction and detection procedures are described more fully in
Pollacco et al. (2006) and Collier Cameron et al. (2006), see also the
companion papers in this series (Christian et al. 2006; Lister et
al. 2007; Street et al. 2007).

% Doublecheck where these units are coming from!!
%\begin{figure}
%\centerline{\hbox{
% %\psfig{file=0516+3126_rmsplot.ps,width=8cm}
% %\psfig{file=0344+2427_rms.ps,width=8cm}
% \psfig{file=rmsplot_resample.ps,width=8cm}
%}}
%\caption{Flux-rms plot for the field 0344+2427 before additional
%  trend-removal through the SysRem algorithm (Tamuz 2005). This field
%  contains lightcurves for 38629 stars from 607 frames over 27
%  nights.}
%\label{rmsplot}
%\end{figure}

\subsection{Photometry Pipeline} 

The collaboration has built a fully-automated data reduction pipeline
that achieves our goal of obtaining photometric precision of $\sim
1\%$ for stars with V $<$ 13. Photometric precision is typically 0.02
mag at $V=13$, with 5 millimag achieved at $V=8.5$. The pipeline uses
custom written f77 programs and several STARLINK packages called from
shell-scripts; it is thus somewhat portable and uses capabilities
already freely available as much as possible. The pipeline itself is
described more fully in Pollacco et al. (2006); here we remark on its
following relevant features: \\

\noindent (1) Frame classification and quality-control is performed on
the input frames through statistical characterisation of the frame
content, with minimal reliance on object headers. Currently $\sim$85\%
of frames are accepted for further processing depending on the
observing conditions during any given night. \\
\noindent (2) Running calibrations are produced by optimally weighting
the calibration history across a season, including exponentially
decreasing weighting with a 14-day timescale to allow for varying
dust-patterns on the lens and other systematics which can vary with
time. This measurably reduces the systematic scatter in the thermal,
flatfield and bias frames. \\
\noindent (3) By triangle-matching selected detected objects with Hipparcos
positions in the Tycho-2 catalogue, a full nine-term plate-solution on
the tangent plane is derived, allowing for pointing errors and
distortion within the glass of the lens by fitting observed stellar
positions directly to their catalogued positions on the sky. \\
\noindent (4) Objects are detected in the frame at $> 4\sigma$ above
background (using a modified version of {\it SExtractor}; Bertin \&
Arnouts 1996). Dedicated f77 routines produce aperture photometry in
three concentric apertures of radius 2.5, 3.5 and 4.5
pixels. Lightcurves using the 3.5-pixel aperture are retained for
further processing; a variant of the curve-of-growth method of Stetson
(1990) is used to affix a blending index to each object based on the
flux evolution with aperture size. \\
\noindent (5) Lightcurves from a given field are processed as an
ensemble to fit the transformation from the instrumental magnitude
system to the Tycho-2 $V$ bandpass. The data are weighted using
inverse variance weights that incorporate, in addition to the formal
errors from the pipeline, variance components that quantify the
intrinsic variability of each star and the patchiness of extinction
across each frame.  These additional variances are estimated using the
maximum-likelihood method described by Collier Cameron et al
(2006). The magnitude zeropoint is determined to a precision of
1-2mmag per frame (Pollacco et al. 2006 and Collier Cameron et
al. 2006). \\
\noindent (6) The photometry is then uploaded to the WASP archive at
Leicester University, which allows rapid access to time-series of
various quantities for each object, through a custom-written
query-language based on SQL. 

% place table 1 here
% put table 1 here!
\noindent \begin{table*}
%\caption{J2000.0 coordinates of field centres surveyed here, showing number of targets suitable for the transit-hunting algorithm, number of objects selected and the number of photometric candidates selected for catalogue-based culling.}
\caption{Field statistics. For each field, we report 1. $N_{cand}$ - the number of targets selected for the BLS search algorithm, 2. $N_{hun}$ - the number of candidates passed forward for visual selection, 3. $N_{vis}$ - the number of candidates passing visual selection (Priority 1 / Priority 2), 4. $N_{S/N}$ - the number of candidates passing further cuts against short periods, ellipsoidal variations and noise signatures (Priority 1 / Priority 2), 5.$N_{f}$ - final number of candidates (Priority 1 / Priority 2). }
\begin{tabular}{ccccccccc}
\hline
RA & Dec & Nights & Frames & $N_{cand}$ & $N_{hun}$  & $N_{vis}$  &  $N_{S/N}$ & $N_{f}$ \\
\hline
0316 & +3126 & 60 & 1882   &   6810 & 162  & 3/5 & 1/0 & 0/1 \\
0317 & +2326 & 60 & 1885   &   5942 & 161  & 3/1 & 3/0 & 0/1 \\
0343 & +3126 & 64 & 1402   &   8465 & 115  & 0/1 & 0/0 & 0/0 \\
0344 & +2427 & 27 & 607    &   6037 & 136  & 0/0 & 0/0 & 0/0 \\
0344 & +3944 & 46 & 1402   &  17615 & 417  & 0/1 & 0/1 & 0/0 \\
0416 & +3126 & 46 & 1400   &  11106 & 231  & 2/1 & 1/0 & 0/0 \\ 
0417 & +2326 & 46 & 1357   &   6241 & 117  & 0/0 & 0/0 & 0/0 \\
0443 & +3126 & 44 & 1029   &   8314 & 147  & 2/1 & 2/0 & 0/0 \\
0444 & +3944 & 43 & 1014   &  20432 & 368  & 0/1 & 0/0 & 0/0 \\
0516 & +3126 & 43 & 1008   &  22406 & 389  & 5/5 & 4/2 & 1/0 \\
0517 & +2326 & 43 & 1013   &  13506 & 219  & 1/5 & 1/4 & 0/1 \\
0543 & +3126 & 37 & 524    &  15021 & 226  & 4/3 & 1/0 & 0/0 \\
\hline
     &      &    & Totals  & 141895 & 2688 & 20/24 & 13/7 & 1/3 \\ 
\hline
\end{tabular}
\label{stats}
\end{table*}

\subsection{Photometric Transit Candidates}

At this stage, small systematic trends are still present in the
photometry; nevertheless, we store these data in the archive rather
than storing detrended data. It was envisaged that detrending routines
would improve over time; this approach thus allows the user to apply
the latest, best routines at the analysis stage. For the work
described here, the generalised linear trend-removal algorithm SYSREM
(Tamuz et al. 2005) was employed to remove the remaining systematic
trends. Investigation is currently underway to fully characterise
these trends for future datasets. Under the nominal observing
strategy, 35 frames per night are taken for well-sampled fields,
however because not all fields are well-sampled under an automated run
(for example a field might have only a few frames taken before dawn),
we cannot assume all objects are well-sampled. Objects were selected
for further analysis for the transit-search if at least 500 points
were recorded over more than 10 nights, with Tycho-2 $V$ $\la 13$.

The resulting set of lightcurves was subjected to automated
application of transit-detection algorithms to isolate the small
subset of transit-candidates. Comparison of the Matched-filter (Street
et al. 2003), Box Least-Squares (Kov\'{a}cs et al. 2002; hereafter
BLS) and Bayesian back-end (Aigrain \& Favata 2002) techniques
suggests BLS is most suited to our purposes (Aigrain \&
Irwin 2004), so was selected as our main transit-search algorithm. Our
own Monte-Carlo simulations of the effectiveness of the transit
algorithms when applied to artificial transits over real noise
lightcurves from WASP, will be reported elsewhere (Enoch et al. in
prep). The BLS algorithm was implemented in a two-stage proces. An
initial coarse-grid search was made over the period-range ($0.9 \le P
\le 5$ days), with the period-range chosen to allow some exploration
of period-space beneath the 1-day boundary, while still producing
well-sampled lightcurves at the long end of the period range. The
period interval is set to ensure distinguishable folded lightcurves,
in the sense that when folded on two successive periods in the
interval, the resulting phase difference of a given feature between
the two lightcurves corresponds to the expected transit width at the
longest period searched. The results of this coarse pass were refined
by a second, finer pass in which the period spacing is now set so that
the phase drift over the entire dataset is less than {\it half } the
expected transit width (Collier Cameron 2006). A typical period
spacing would thus be 0.002d for the coarse search and 0.001d for the
finer search. For each candidate, fit statistics and parameters of the
best-fit transit model at this stage were obtained for the five most
significant period detections. Detections were ranked by the fit
statistic $\Delta \chi^2$, which gives the improvement of the
best-fitting transit-model over a flat lightcurve model, and is our
adopted proxy for the transit S/N detection.

\begin{figure}
\begin{center}
\psfig{file=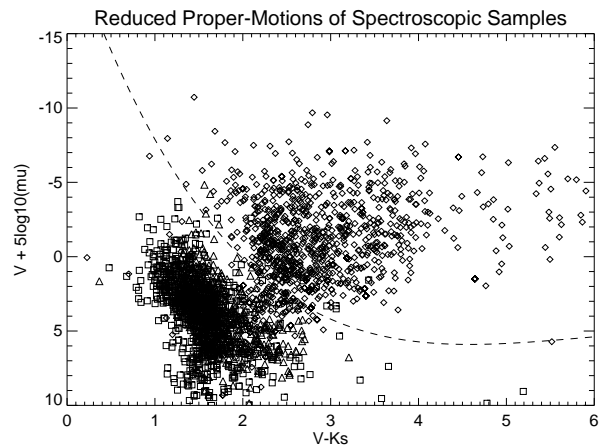,width=8.5cm}
\caption{Reduced Proper-motion vs ($V-K_S$) colour of a selection of
  giants and dwarfs from the Cayrel et al (2001) and Valenti \&
  Fischer (2005) surveys (using proper motions in mas
  yr$^{-1}$). Diamonds: Cayrel et al. giants. Boxes: Cayrel et
  al. dwarfs. Triangles: Valenti \& Fischer dwarfs. The dashed line
  shows a polynomial boundary (as a function of $V_{SW}$-$K_S$)
  constructed to discriminate between the two
  regions. This boundary serves as a guide for automatic
  classification, however the position of each photometric
  transit-candidate was visually checked in this diagram for
  assessment of luminosity class.}
\end{center}
\label{fig:rpm}
\end{figure}

Filtering of the candidate-list was then applied based on: (i)
repetition of a transit-like event, (ii) reduced chi-squared statistic
of the best-fit transit model $\chi^2_{\nu} < 3.5$, (iii) the presence
of any gaps in the folded transit lightcurve a factor $>$ 2.5 longer
than the transit width - indicative of a fit dominated by sampling
gaps, (iv) the signal to noise ratio in the presence of correlated
noise (commonly called ``red-noise'') $S_{red}$ (Pont et al. 2006),
and (v) the transit to anti-transit ratio $\Delta \chi^2$/$\Delta
\chi^2_{-}$. The latter measures the improvement in fit-statistic
$\Delta \chi^2$ when a transit model consisting of regular intensity
dips is fit, scaled by the improvement $\Delta \chi^2_{-}$ when an
''anti-transit'' model consisting of regular flux {\it brightenings}
is fit instead. This statistic can be used to characterise lightcurves
with a strong correlated noise component (Burke et al. 2006).

Correlated noise introduces significant systematics which raise the
detection threshold for significant periodicities in time-series
data. Although well-characterized in several fields in astrophysics
(such as X-ray variability studies; e.g. Homer et al. 2001), its
relevance to optical searches for exoplanets was not fully appreciated
when the ground-based transit surveys were planned, and thus deserves
some amplification here (see also the Discussion, Smith et al. 2006
and Collier Cameron et al. 2006). For ground-based photometric surveys
the errors in measurement are usually correlated on timescales of tens
of minutes to hours, producing a low-frequency component to the noise
that can mimic an exoplanet transit. $N_{tr}$ transits are observed,
with $L_i$ measurements in each transit. The transit is assumed to be
a step-function to the photometric precision of SW-N so that each
datapoint observed during transit is treated as an estimate of the
full transit depth $\delta$. These estimates are binned by transit
number, with corresponding binned measurement error
$\sigma_{bin,i}$. The signal to noise estimate for the full set of
transit measurements is then
\begin{equation}
  S^2_{red}  = \sum_{i=1}^{N_{tr}} \frac{\delta_i^2}{\sigma^2_{bin,i}(L_i)}
\end{equation}
\noindent We require a prescription for the relation between binned
measurement error $\sigma_{bin}$ and unbinned error $\sigma_u$ in the
presence of real noise. In the case of SW-N data, this relationship is
characterised as
\begin{equation}
  \sigma_{bin} = \sigma_{u}L^b
\label{eqn_2}
\end{equation}
\noindent Pure uncorrelated noise would show the familiar $b=-0.5$
while binning would not improve matters for fully correlated noise and
thus $b=0$. For each star, out-of-transit data from each night is used
estimate the index $b$ from a fit to $\sigma_{bin}/\sigma_u$ as a
function of $L$. The relation (\ref{eqn_2}) is then used to relate the
unbinned rms scatter observed during transit $\sigma_{u,i}$ to the rms
scatter of the binned estimate of the transit depth. In reality, $L_i$
and $b$ will vary on a night-by-night basis; in order to filter on a
star-by-star basis we take the average across values $L$ and $b$
across the observed transits, leading to
\begin{equation}
  S_{red} = \frac{\delta \sqrt{N_t}}{\sigma_u L^b}
\end{equation}
\noindent where the rms scatter of the unbinned data $\sigma_u$ is now
taken across all the datapoints during transit. Further information is
given in Collier Cameron et al. (2006); note that for objects with $V
< 11$ the covariance parameter $b$ is not quite -0.5 after detrending,
which suggests a low level of residual structure may be present in the
detrended lightcurves.

We remind the reader that we are ranking periodicities from each
candidate by several criteria in the same search, so care must be
taken to interpret the ranking that results. In two cases, the attempt
by the algorithm to maximize signal to noise in the presence of
correlated noise $S_{red}$ caused the returned best-fit period to jump
to a shorter period that was much less significant than the most
significant trough in the BLS periodogram. The $S_{red}$ statistic can
become very low at certain pathological frequencies which beat with
the day-night cycle, producing a much higher-than-average number of
observable transits. Even if the $\Delta\chi^2$ is not highly
significant at such frequencies, $S_{red}$ can thus become very
large. $S_{red}$ is therefore only used to determine whether the
frequencies associated with the strongest $\Delta\chi2$ actually yield
significant detections when the contribution from correlated noise is
considered.

%SW-N typically samples
%$\sim$40\% of any given transit-like event. When evaluating the
%S$_{red}$ of the putative transit signal at a trial period, the
%lightcurve is folded on this trial period, with the in-transit
%phase-range determined from the mean folded lightcurve. Including
%correlated noise as well as genuine transits, the identification of
%dipping events with the mean transit thus depends on the trial period
%used. For pathologically short periods, an artificially high number of
%transits can be mistakenly identified, leading to an artificially low
%S$_{red}$. Such a case is usually rather straightforward to identify
%due to a sudden increase in the number of transits observed in a given
%trial period compared to the other promising periods.

Detections with the five highest S$_{red}$ values are produced for
each candidates; where the best detections show similar $S_{red}$ we
retained the detection corresponding to the most significant trough in
the BLS periodogram (for example the candidate
J025922.67+275416.0). In one case the most significant trough in the
BLS periodogram only sampled two transits, so we rejected that period
and chose the second strongest (J051849.56+211513.6). In most cases,
the most significant detection in the BLS periodogram was clearly much
more significant than its nearest rivals (see Figure 2) however in at
least one case several marginally less significant period-detections
were also reported (J051849.56+211513.6; Section 4); no account was
taken of these secondary detections in this case.

The result is a set of 2688 transit-candidates ranked by the
fit-statistic $\Delta \chi^2$.
%, which gives the improvement of the
%best-fitting transit-model over a flat lightcurve model, and is our
%adopted proxy for the transit S/N detection.
The lightcurves and BLS periodograms for each of these objects were
visually examined, to remove lightcurves dominated by obvious sampling
effects and other artefacts. This examination was carried out
independently by the first two authors and the final list produced
after comparison of the analyses. Objects are deselected from further
consideration if their lightcurves meet any two of the following
criteria:
%\footnote[3]{These criteria were deliberately set somewhat generously
%  to avoid rejection of genuine transit-events.}:

\noindent 1. Folded lightcurve dominated by sampling gaps. \\
2. Most significantly detected period and the nearest alias of the 1-d sampling are indistinguishable from each other in the BLS periodogram. \\
3. Visible out-of-transit variability both above and below the mean flux level. \\
4. Photometric transit-depth, $\delta$, greater than $15 \%$. \\
% (see Figure \ref{depthrad})\\
5. Deep, V-shaped lightcurve suggestive of stellar transit. \footnote{This criterion was used for objects with $\ga 10\%$ intensity dips that are clearly stellar binaries; more marginal cases are retained as possible exoplanet candidates.}  \\
6. Ellipsoidal trends apparent in folded lightcurve. \\
7. Multiple transit events are apparent in the folded lightcurve, suggesting an incorrect period has been used, and the corrected period is outside the 0.9-5 day period range. \\
8. Transit duration greater than 5 hours. \\
9. Only two apparent transit events present in the entire lightcurve (if a candidate meets this criteria it is removed from further consideration).\\

% THIS FIGURE HAS BEEN REMOVED ENTIRELY.
%\begin{figure}
% \psfig{file=depths.ps,width=8cm,angle=-90}
%\caption{Transit depth as a function of (H-K) colour for planet radii 0.5 R$_J$ (solid line), 1.0 R$_J$ (dashed line) and 1.5 R$_J$ (dot-dashed line). Filled asterisk - Priority 1, open circles - Priority 2. The stellar spectral type is computed as a function of ({\it H-K}) color for main-sequence objects following the Bessell \& Brett (1988) relation (after transforming the colour from 2MASS to the appropriate system; see also Glass 1999).}
%\label{depthrad}
%\end{figure}
%\begin{figure}
%\begin{center}
%\psfig{file=Classification.ps,width=8.5cm}
%\caption{Reduced Proper-motion vs ($V-K_S$) colour of a selection of
%  giants and dwarfs from the Cayrel et al (2001) and Valenti \&
%  Fischer (2005) surveys (using proper motions in mas
%  yr$^{-1}$). Diamonds: Cayrel et al. giants. Boxes: Cayrel et
%  al. dwarfs. Triangles: Valenti \& Fischer dwarfs. The dashed line
%  shows a polynomial boundary (as a function of $V_{SW}$-$K_S$)
%  constructed by one of us (DMW) to discriminate between the two
%  regions. This boundary serves as a guide for automatic
%  classification, however the position of each photometric
%  transit-candidate was visually checked in this diagram for
%  assessment of luminosity class.}
%\end{center}
%\label{fig:rpm}
%\end{figure}

This visual inspection trimmed the 2688 candidates further to 44,
comprising 20 targets considered likely from the photometry to contain
a transiting extrasolar planet (Priority 1) and 24 candidates where
just one of the above tests are failed by the candidates (Priority 2
candidates). At this stage, a number of cuts were made on the
surviving objects based on lightcurve statistics returned from the
period-search and lightcurve analysis. Objects were only passed
forward as candidates if:

\noindent 1. S/N in the presence of correlated noise $S_{red} >8$ (c.f. Pont
et al. 2006). \\
\noindent 2. Period $\geq$ 1.05 days. \\
\noindent 3. S/N of ellipsoidal variations $<$ 8.0 (c.f. Sirko \& Paczy\'{n}ski (2003). \\
\noindent 4. Transit to anti-transit ratio $\Delta \chi^2/ \Delta \chi^2_{-} \ge 2.0$ (c.f. Burke et al. 2006). \\

All but 20 of the remaining candidates were filtered out by these
steps. In summary, then, a typical field would contain several hundred
raw candidates, out of which visual inspection would leave 1-2
Priority 1 and 3-4 Priority 2 candidates; however further cuts against
correlated noise, ellipsoidal variations and period would reduce this
number by about half (see Table \ref{stats}).

\begin{table*}
\caption{Candidates from the BLS search that pass initial visual
  inspection. N$_{tr}$ denotes the number of transits observed, $n_t$
  the number of valid observations of the object, $n$ the number of
  valid points during transit $\Delta \chi^2$ the improvement of the
  best-fitting transit-model over a flat lightcurve model,
  $\Delta \chi^2$/$\Delta \chi^2_{-}$ the ratio of this fitting
  statistic when using the transit model to an ``anti-transit''
  brightening model (Section 3.3), S$_{ell}$ the signal to noise of
  ellipsoidal variation, S$_{red}$ the signal to noise including
  correlated noise. The final two columns give the priority accorded
  the candidate at the stage of visual examination and the primary
  reason for its rejection (if applicable). Reasons for removal are:
  low S/N against correlated ``red''-noise (R), presence of
  ellipsoidal variations (E) and low $\Delta \chi^2$/$\Delta \chi^2_{-}$ (A).}
\begin{tabular}{lcccccccccccl}
\hline
SWASP ID & Period & Duration & Depth & N$_{tr}$ & $n_{pts}$ & $n$ & $\Delta \chi^2$ & $\frac{\Delta \chi^2}{\Delta \chi^2_{-}}$ & S$_{ell}$ & S$_{red}$ & Vis & Cut \\
         & (d)    &  (hrs)   & (mag) & & & & & & & & \\
\hline
J025922.67+275416.0 & 1.098797 & 2.38 & 0.0179 &  8 & 1693 & 129 & 154.226 & 2.734 & 1.051 & 6.923 & P2 & R \\ % 2592
J025947.03+283310.4 & 3.074289 & 4.61 & 0.0132 &  5 & 1694 & 80 & 219.119 & 2.826 & 3.298 & 7.335 & P2 & R \\ % 1170
%SW0316+3126 & DAS3 & 0 & P2 \\                    
J030117.53+274943.0 & 3.070961 & 3.86 & 0.0241 &  4 & 1693 & 64 & 103.562 & 1.661 & 1.416 & 6.936 & P2 & R \\ % 3705
J030153.95+332213.0 & 2.350089 & 5.18 & 0.0450 & 12 & 1694 & 228 & 660.272 & 4.386 & 0.399 & 7.321 & P2 & R \\ % 6164
J031632.80+300144.2 & 2.198882 & 3.29 & 0.0310 &  6 & 1692 & 112 & 1457.156 & 5.896 & 25.784 & 14.364 & P1 & E \\ % 505
J032515.17+341031.6 & 1.011542 & 3.29 & 0.0374 & 17 & 1693 & 300 & 582.241 & 4.481 & 6.685 & 11.121 & P1 & P \\ % 5485
J032739.88+305511.3 & 1.051158 & 3.00 & 0.0253 & 14 & 1693 & 234 & 911.154 & 5.764 & 2.966 & 9.758 & P1 & \\ % 3326 
J033503.83+325915.2 & 2.135410 & 2.33 & 0.0650 &  9 & 1694 & 106 & 929.219 & 2.793 & 15.031 & 9.837 & P2 & E \\ %4626 
\hline                           
J030157.61+204037.1 & 1.571295 & 3.17 & 0.0807 &  9 & 1690 & 144 & 2213.008 & 37.633 & 3.246 & 12.431 & P1 & \\ % 3330 
J030854.44+234517.4 & 2.206365 & 4.25 & 0.0567 &  7 & 1434 & 101 & 616.065 & 8.403 & 6.469 & 16.135 & P1 & \\ % 2728 
J031103.19+211141.4 & 2.730148 & 3.46 & 0.0403 &  5 & 1690 & 89 & 712.650 & 12.847 & 5.348 & 9.077 & P1 & \\ % 3119 
J033042.00+243027.9 & 3.178541 & 3.98 & 0.0157 &  3 & 1690 & 72 & 613.552 & 10.072 & 11.575 & 6.354 & P2 & R \\ % 433
\hline                           
J034747.35+350105.7 & 1.928731 & 2.16 & 0.0254 &  5 & 1267 & 163 & 2847.39 & 11.321 & 13.692 & 7.533 & P2 & R \\ % 52
\hline                           
J034628.00+365747.0 & 1.856870 & 2.33 & 0.0703 &  3 & 1242 & 63 & 736.339 & 16.640 & 0.877 & 8.105 & P2 & \\ %11093  
\hline                           
J041411.76+302105.0 & 2.554799 & 4.58 & 0.0565 &  6 & 1312 & 115 & 1349.406 & 17.297 & 6.050 & 9.292 & P1 & \\ % R 6536
J042255.90+290701.5 & 2.054940 & 1.80 & 0.0680 &  6 & 1312 & 138 & 2083.12 & 6.685 & 29.186 & 8.669 & P2 & E \\ % 4599
J042518.63+305018.1 & 1.265071 & 1.78 & 0.1036 &  8 & 1312 & 83 & 6393.781 & 9.069 & 11.243 & 10.328 & P1 & E \\ % 3564
\hline                           
J045349.66+333842.5 & 1.843365 & 4.27 & 0.0340 &  6 & 913 & 94 & 422.898 & 1.694 & 11.032 & 8.736 & P2 & A \\ % 2892
J045441.00+335323.2 & 1.435404 & 1.97 & 0.1118 &  5 & 913 & 45 & 871.943 & 10.557 & 2.699 & 11.698 & P1 & \\ % 6521 
J044803.38+342415.5 & 1.385160 & 3.10 & 0.1202 &  7 & 913 & 102 & 1718.045 & 15.560 & 4.113 & 10.556 & P1 & \\ % 7340 
\hline                           
J050328.03+394509.4 & 1.727674 & 2.16 & 0.0431 &  3 & 619 & 61 & 796.254 & 7.185 & 15.228 & 5.132 & P2 & R \\ % 2784
\hline                           
J050712.55+335934.4 & 1.389950 & 2.09 & 0.0195 &  6 & 826 & 72 & 351.477 & 7.992 & 1.392 & 8.023 & P1 & \\ % 2671 
J050917.50+300309.8 & 1.923790 & 1.78 & 0.0274 &  5 & 826 & 114 & 373.920 & 2.815 & 7.905 & 6.973 & P2 & R \\ %6374 
%SW0516+3126 & DAS3 & 0 & P2 \\                    
J051221.34+300634.9 & 1.237851 & 1.87 & 0.0304 &  5 & 822 & 49 & 977.310 & 15.529 & 0.125 & 9.080 & P1 & \\ % 2142 
J051414.50+350639.9 & 1.659918 & 2.47 & 0.1866 &  8 & 816 & 122 & 2770.867 & 14.260 & 12.015 & 9.692 & P1 & E \\ % 22023
J051632.17+304921.5 & 2.558843 & 5.83 & 0.0537 &  8 & 824 & 102 & 404.012 & 14.722 & 4.356 & 9.510 & P1 & \\ % 20005 
J052123.50+343759.3 & 1.911629 & 2.14 & 0.1091 &  6 & 826 & 32 & 822.981 & 8.824 & 3.169 & 12.427 & P2 & \\ % 13753 
J052155.26+334037.0 & 1.820449 & 2.35 & 0.0255 &  3 & 825 & 43 & 274.351 & 17.576 & 0.377 & 7.871 & P2 & R \\ % 4884
J052155.29+311153.2 & 2.552743 & 2.33 & 0.0172 &  4 & 823 & 46 & 112.207 & 8.394 & 0.599 & 10.579 & P2 & \\ %2360  
J052639.24+341813.9 & 1.172678 & 2.78 & 0.0843 &  6 & 826 & 74 & 4831.498 & 60.055 & 17.855 & 9.546 & P2 & E \\ %4054 
J053442.52+312922.3 & 1.675041 & 2.40 & 0.0857 &  3 & 826 & 38 & 2048.602 & 24.867 & 0.544 & 9.203 & P1 & \\ % 4526 
\hline 
J050210.19+222523.8 & 1.968182 & 2.78 & 0.0968 &  4 & 834 & 38 & 1338.312 & 7.930 & 5.896 & 12.496 & P2 & \\ % 2787 
J050241.49+235554.6 & 4.148943 & 5.26 & 0.0753 &  3 & 834 & 40 & 755.581 & 6.135 & 1.365 & 9.933 & P1 & \\ % 2576 
J050642.37+214850.2 & 1.620502 & 2.54 & 0.0857 &  3 & 834 & 35 & 1264.624 & 33.940 & 3.185 & 9.215 & P2 & \\ % P 4952
J051108.55+230632.3 & 1.709274 & 2.59 & 0.0235 &  4 & 819 & 66 & 948.96 & 2.809 & 22.846 & 7.396 & P2 & E \\ %213 
J051109.87+222428.3 & 1.391621 & 2.90 & 0.0306 &  5 & 834 & 68 & 310.172 & 3.820 & 3.518 & 8.258 & P2 & \\ % 3497 
J051849.56+211513.6 & 1.348566 & 2.28 & 0.0579 &  6 & 834 & 57 & 211.862 & 4.363 & 1.103 & 13.136 & P2 & \\ % 5344 
\hline                         
J053026.87+350839.4 & 1.225148 & 2.16 & 0.0668 &  5 & 524 & 50 & 512.587 & 20.940 & 5.471 & 7.767 & P1 & R \\ % 9287
J053428.54+331646.7 & 1.227779 & 4.99 & 0.0672 &  6 & 523 & 86 & 1357.361 & 19.554 & 0.455 & 7.339 & P2 & R \\ %rej10740 
J053430.23+331610.6 & 1.229169 & 4.73 & 0.0393 &  7 & 523 & 84 & 759.293 & 8.590 & 0.170 & 7.700 & P1 & R \\ % 7472
J054511.65+323330.7 & 1.553595 & 1.49 & 0.0486 &  3 & 519 & 20 & 507.882 & 9.426 & 1.155 & 11.655 & P1 & \\ % 2146 
J054645.34+292753.7 & 1.175271 & 2.11 & 0.0620 &  3 & 512 & 31 & 414.517 & 1.839 & 3.775 & 13.211 & P2 & A \\ % 915
J055303.05+275339.4 & 2.410921 & 3.34 & 0.0628 &  3 & 524 & 37 & 1552.535 & 15.285 & 4.518 & 6.238 & P1 & R \\ % 6759
J055557.92+283738.4 & 1.241766 & 2.66 & 0.0198 &  5 & 521 & 51 & 105.570 & 1.543 & 1.690 & 8.076 & P2 & A \\ % 5720
\hline
\end{tabular}
%\caption{Candidates from the BLS search that pass initial visual
%  inspection. 27/45 are removed from further consideration through the
%  cuts described in Section 3.2. Reasons for removal are: low S/N
%  against correlated noise (R), short period (P), presence of ellipsoidal
%  variations (E) and low ratio of $\Delta \chi^2$ using transit to
%  anti-transit models (A).}
\label{cuts}
\end{table*}

\subsection{Catalogue-based Assessment}

The final cut is the use of prior knowledge from previous surveys with
higher spatial resolution and multi-filter information to remove
surviving systems that are likely to be blends or other astrophysical
false-positives. This stage cut the list of candidates still further.

As we remarked in section 2.2, the depth of the 2MASS, Tycho-2 and
USNO surveys allow the suitability of the remaining candidates to be
assessed on the basis of their colours and proximity to potential
photometric crowding objects. A custom-built online query-tool was
implemented by the Consortium to query a variety of astronomical
catalogues at the position of the transit-candidates, returning survey
images of the target field and multiwavelength information for the
target and nearby objects from which the parent-stellar parameters can
be estimated. We refer the reader to Wilson et al. (2006) for more
detailed information on this process.

{\it Luminosity Class and Spectral Type:} As pointed out by Gould \&
Morgan (2003), roughly 90\% of the bright stars surveyed by
ground-based exoplanet transit-searches, are giants for which a
transiting exoplanet would produce well under 1\% dips; this predicts
a rather high astrophysical false-positive rate (Brown 2003). Stellar
populations in the galactic disk show coherent, restricted velocity
distributions (e.g. Binney \& Merrifield 1998). The reduced proper
motion (hereafter RPM) can be used to kinematically segregate members
of nearby stellar populations; in particular its correlation with
absolute magnitude allows WASP targets with proper motions (available
from the USNO catalogue for most objects) to be roughly classified by
Luminosity class. The position of the target in \{RPM,$(V-K_S)$\}
space is determined using the Tycho-2 V-magnitude (using the observed
SuperWASP V-magnitude $V_{SW}$ as a check), catalogue $K_S$ and proper
motion estimates from the USNO catalogue. The luminosity class
division is based upon spectroscopic surveys of a number of nearby
objects, in particular the Cayrel et al (2001) catalogue and the
Valenti \& Fischer (2005) catalogue from the N2K survey, with the
luminosity class estimated with reference to this fit (Figure
1). Unknown reddening is in principle a systematic bias with this
measure, as it causes the observed ($V-K_S$) and RPM to both be
artificially higher than the intrinsic properties.
The reduced proper motion diagnostic was checked manually for cases in
which it was in any way ambiguous; in particular, the
\{RPM,(V-$K_S$)\} diagnostic becomes somewhat inconclusive for objects
with {\it both} $ 1.5 \la (V-K_S) \la 2.2$ and {\it RPM} $\la $2
(Figure 1). Estimates of the astrometric accuracy of stellar positions
on the plates used, combined with transformation errors, produce a
quality flag in USNO that gives a probability estimate of the reported
proper-motion being correct (Monet et al 2003). Low values of this
quality flag suggest poor proper motion measurement. We also use the
flux-angular diameter relation from interferometric studies (Kervella
et al. 2004, Foqu\'{e} \& Gieren 1997); given catalogue $B,K_S$
magnitudes the angular diameter can be inferred and converted into
stellar radius using the Hipparcos parallax (if available) to infer
distance. If the available information is ambiguous as to the
luminosity class of a target, it is reduced in priority.
%We thus use
%two additional estimates for the luminosity class of the targets. For
%intrinsic colours redder than {\it (H-K$_s$)} $\sim$0.13,
%main-sequence objects start to separate from giants in the {\it
%  (J-H)/(H-K$_S$)} diagram, corresponding to spectral types K7 and
%later (Bessell \& Brett 1988; Glass 1999). Thus for sufficiently
%late-type objects, near-IR colors can be used to estimate luminosity
%class; we use the Carpenter (2001) transformations to convert {\it
%  2MASS} colours into the Bessell \& Brett (1988) system (close to the
%SAAO system; Carter 1990) to make this comparison. For late-type stars
%the Carpenter transformations are unreliable due to differences in
%{\it 2MASS K$_S$} and Bessell \& Brett $K$ filter-transmissions; we
%thus use Carpenter-transformed {\it (J-H)} as a cross-check for
%objects at K0 or later. 

%criteria below were met. Candidates meeting only one of the criteria
%below were demoted to Priority-2 follow-up targets. \\

%\noindent 1. Any object brighter than the candidate within the 48'' SW-N
%aperture 
%\\ 2. Any object $< 5$ mags fainter than the candidate in Tycho-2 $V$
%or USNO-$R$ (one demotion per blending object) 
%\\ 3. Candidate is a giant.
%\\ 4. Candidate is hotter than $F0$ spectral type. 
%\\ 5. Tingley \& Sackett $$\eta$$ outside the 0.6-1.5 range \\

%In practice, no candidates were rejected on criterion 5 alone. In
%addition, candidates were further removed from consideration if they
%coincided with previously-known objects with peculiarities suggestive
%of unusual system components (these objects are listed in section 4.3
%for reference).

{\it Stellar Radius:} For main-sequence stars the transit depth
combined with the 2MASS ($J$-$H$) colour also provides an estimate for
the spectral type and radius of the parent star, and thus the radius
of the putative planet (Ammons et al. 2006). Note that
because Ammons et al. (2006) also used 2MASS photometry,
transformation from 2MASS into e.g. the Bessell \& Brett (1988) system
is not required to estimate radii, removing a potential source of
systematic error. For dwarfs the ($J$-$H$)-radius relationship suffers
from a degeneracy, in that the relationship turns over at spectral
type $\sim$M$0$ (Bessell \& Brett 1988). As a cross-check, we use
($V$-$K_S$) and ($B$-$V$) to estimate effective temperature (Blackwell
\& Lynas-Gray 1994; Zombeck 1992), and the radius from the standard
temperature-radius relation for main-sequence stars (Gray
1992). $B$-magnitudes for this step are taken from USNO, $V$ from
Tycho 2 or if unavailable, from SW-N. The observed transit-depth is
used to estimate the planetary radius assuming the occultation is due
to the full disk of the planet at maximum depth. ($V$-$K_S$) colors
also provide a way to break the ($J$-$H$)-radius degeneracy; should an
object show observed ($V$-$K_S$) too blue for a M0 dwarf, the
($J$-$H$) color must correspond to spectral type earlier than M$0$ -
in practice this applies for all the candidates presented here.

\begin{table*}
\caption{Lightcurve timing information for the candidates. HJD of mid-transit = 2450000.0 + Epoch. $\eta$ is the Tingley \& Sackett figure of merit for identification with a transiting exoplanet (Tingley \& Sackett 2005). Errors here and throughout this report are formal 1$\sigma$ errors on transit model fits to the data.}
\begin{tabular}{cccccccc}
\hline
SWASP ID & Epoch (d) & Period (d) & Duration (h) & Depth (\%) & $\eta$ & N$_{trans}$ & V$_{SW}$\\
\hline
%1SWASP J031103.19+211141.4  & 3193.2342 $\pm 0.0023$ & 2.7301 $\pm 1.03 \times 10^{-4}$ & 3.46 $\pm 0.12$ & 4.03 $\pm 0.14$ & 6 & 12.23 \\
%1SWASP J032739.88+305511.3  & 3194.3954 $\pm 0.0014$ & 1.0512 $\pm 2.60 \times 10^{-5}$ & 3.00 $\pm 0.07$ & 2.53 $\pm 0.08$ & 14 & 12.17 \\
%1SWASP J031632.80+300144.2  & P1 THIS OBJECT NO LONGER A CANDIDATE
% have corrected these coords from becky's table
1SWASP J051221.34+300634.9 & 3218.6880 $\pm 0.0009$  & 1.2379 $\pm 2.7 \times 10^{-5}$ & 1.872 $\pm 0.048$ & 3.04 $\pm 0.09$ & 0.77 & 5 & 10.90 \\
\hline
1SWASP J031103.19+211141.4  & 3193.2342 $\pm 0.0023$ & 2.7301 $\pm 1.03 \times 10^{-4}$ & 3.46 $\pm 0.12$ & 4.03 $\pm 0.14$ & 0.94 & 5 & 12.23 \\
1SWASP J032739.88+305511.3  & 3194.3954 $\pm 0.0014$ & 1.0512 $\pm 2.60 \times 10^{-5}$ & 3.00 $\pm 0.07$ & 2.53 $\pm 0.08$ & 1.06 & 14 & 12.17 \\
1SWASP J051849.56+211513.6 &  3219.3183 $\pm 0.0024$ & 1.3486 $\pm 6.20 \times 10^{-5}$& 2.28 $\pm 0.12$  & 5.79 $\pm 0.39$ & 0.88 & 6 & 12.05 \\
%1SWASP J030021.76+275654.3 &  P2 & 3.0577 & 4.5120 & 1.67 & 6 & 10.9196 \\
% the following object is rejected at the signal-to-correlated noise cut!!
%1SWASP J025922.67+275416.0 & 3193.6806 $\pm 0.0034$ & 2.7404 $\pm 1.67 \times 10^{-4}$ & 2.304 $\pm 0.17$ & 2.41 $\pm 0.19$  & 5 & 11.90 \\
%1SWASP J051849.56+211513.6   P2 \\
\hline
\end{tabular}
%\caption{Lightcurve timing information for the candidates. HJD of mid-transit = 2450000.0 + Epoch}
\label{cands}
\end{table*}

\begin{figure*}
%  \centerline{\hbox{
%  \psfig{file=J031103.19+211141.4_lc.ps,width=8cm}
%  \psfig{file=J031103.19+211141.4_pd.ps,width=8cm}
%  }}
%\vspace{2mm}
%  \centerline{\hbox{
%  \psfig{file=J032739.88+305513.3_lc.ps,width=8cm}
%  \psfig{file=J032739.88+305513.3_pd.ps,width=8cm}
%  }}
\vspace{2mm}
  \centerline{\hbox{
  \psfig{file=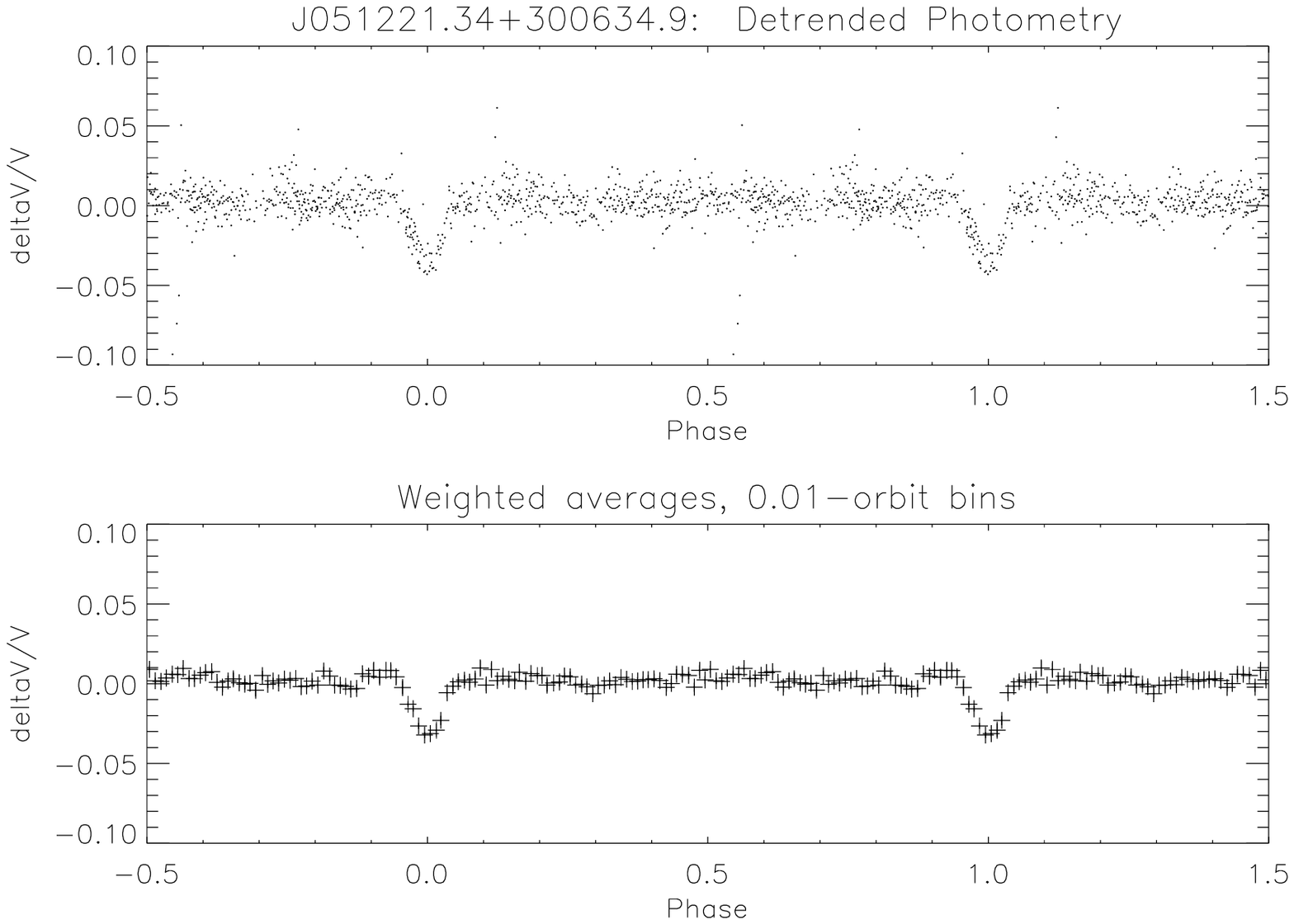,width=8cm}
  \psfig{file=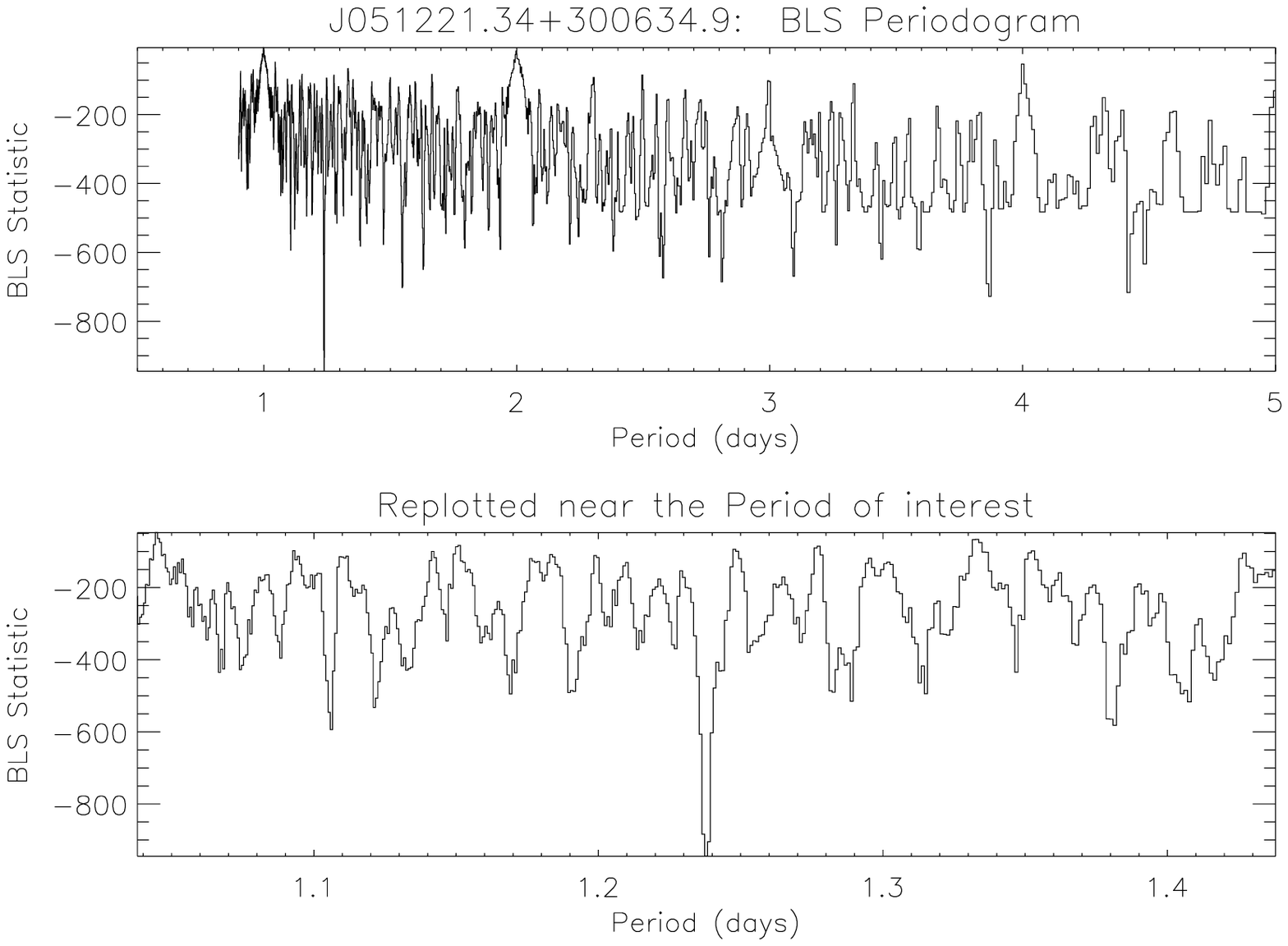,width=8cm}
  }}
% this object has been moved to the rank-two objects
%\vspace{2mm}
%  \centerline{\hbox{
%  \psfig{file=J051849.56+211513.6_lc.ps,width=8cm}
%  \psfig{file=J051849.56+211513.6_pd.ps,width=8cm}
%  }}
  \caption{The accepted Priority 1 Candidate J051221.34+300634.9. {\it
      Left:} Folded lightcurve. Top panel: folded lightcurve after
    detrending. Bottom panel: phase-binned averages weighted by
    $(1/\sigma^2_i)$, where $\sigma^2_i$ is the estimated variance on
    each datapoint including both formal and systematic error (section
    3.1). {\it Right:} Box Least Squares periodogram.}
\label{pri1fig}
\end{figure*}

SW-N routinely achieves photometric precision $\sim 5$mmag at $V=8.5$,
rising to $0.02$ mag at $V=13$; Tycho-2 shows photometric error $\sim
0.05$ mag at V=10.0-11.0, rising to 0.11 mag at V=11.0-12.0 (H$\o$g et
al. 2000). 2MASS observations of calibration standards show rms
residuals of order $\ga$0.05 mag in the $(10 \le H \le 14)$ range
(Nikolaev et al. 2000, Carpenter 2001), so we may expect photometric
errors to be comparable to reddening effects for comparatively high
reddening. For example, with absolute magnitude $M_V \sim 4$, a
typical late-F / early G-dwarf located roughly 200pc from the Sun
would be measured at Tycho-2 $V\sim10.5$ mag. For the fields of
interest here, the local HI column density out to this distance is of
order 10$^{20}$ cm$^{-2}$ (Fruscione et al. 1994), leading to
reddening E(B-V)$\sim$0.02 and extinction $A_V \sim$0.06 (c.f. Binney
\& Merrifield 1998). Thus a subset of objects in the survey will show
uncertain extinction in $V$ that is comparable to the photometric
uncertainty associated with $(V-K_S)$. We thus use parameters inferred
from the 2MASS ($J$-$H$) color preferentially over ($V$ - $K_S$) when
the two measures disagree.

The most inflated planet currently known has radius $R\sim
1.44 R_J$ (Charbonneau et al. 2007a), so we regard SW-N candidates with
inferred radii $\la 1.5R_J$ as sensible candidates. However we do not
reject outright candidates with slightly larger inferred radii to
allow for photometric uncertainty in this detection survey.

%%J042512.91+292308.0 & 10.043 & 1307 & 99.684  & 2.234  & 0.0065 & 1.848 &  4 & 1.09 & 0.75 & 0.69 & 1.49 & 0.21 & F2V   & 2 \\
%\end{tabular}
%\caption{Priority-1 Candidates for follow-up ordered in decreasing $S/N$}
%\end{table*}

% and Tingley \&
%Sackett $$\eta$ \sim$0.84.

\begin{figure*}
%  \centerline{\hbox{
%      \psfig{file=J030021.76+275654.3_lc.ps,width=8cm}
%      \psfig{file=J030021.76+275654.3_pd.ps,width=8cm} 
%  }}
\vspace{2mm}
  \centerline{\hbox{
  \psfig{file=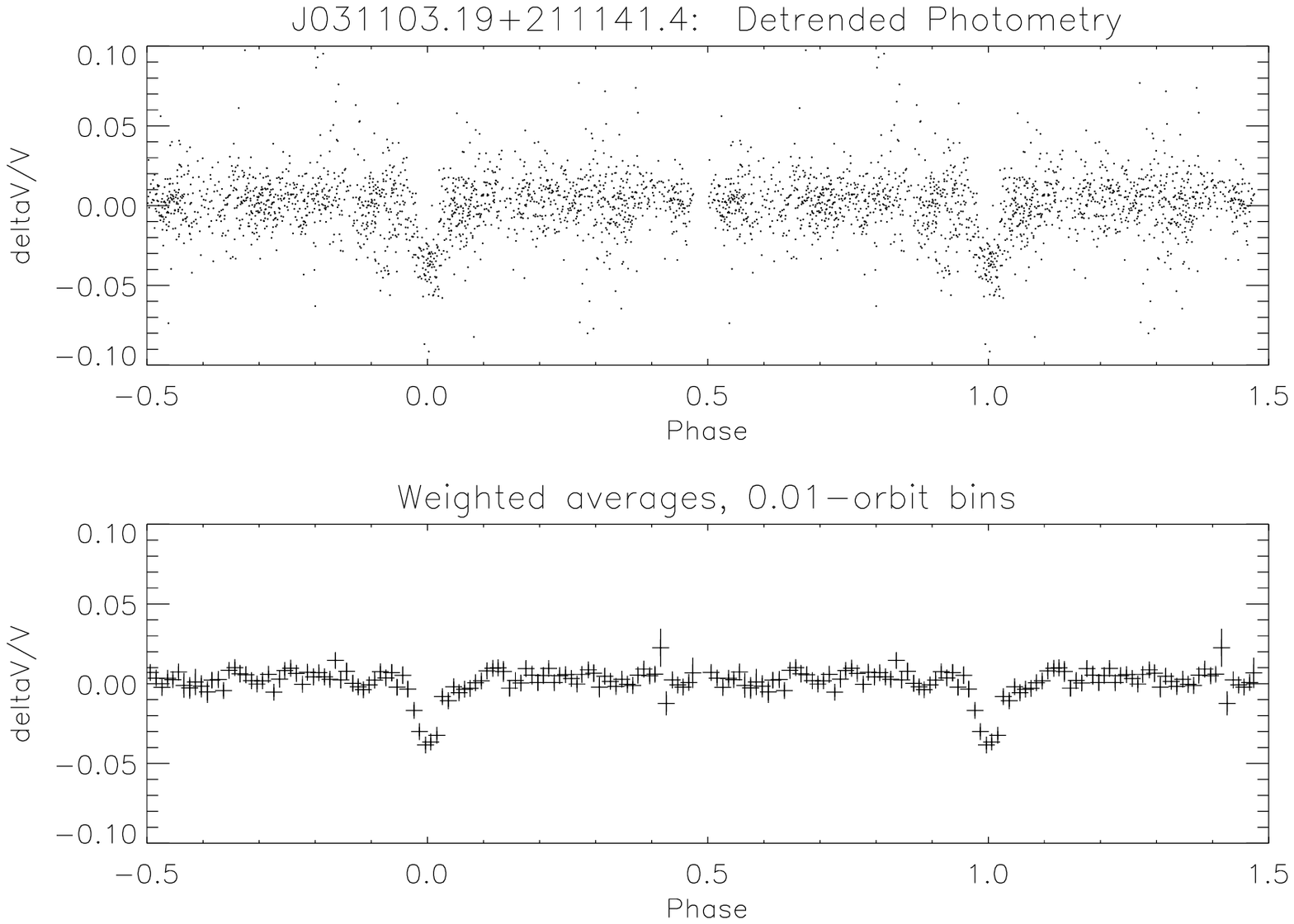,width=8cm}
  \psfig{file=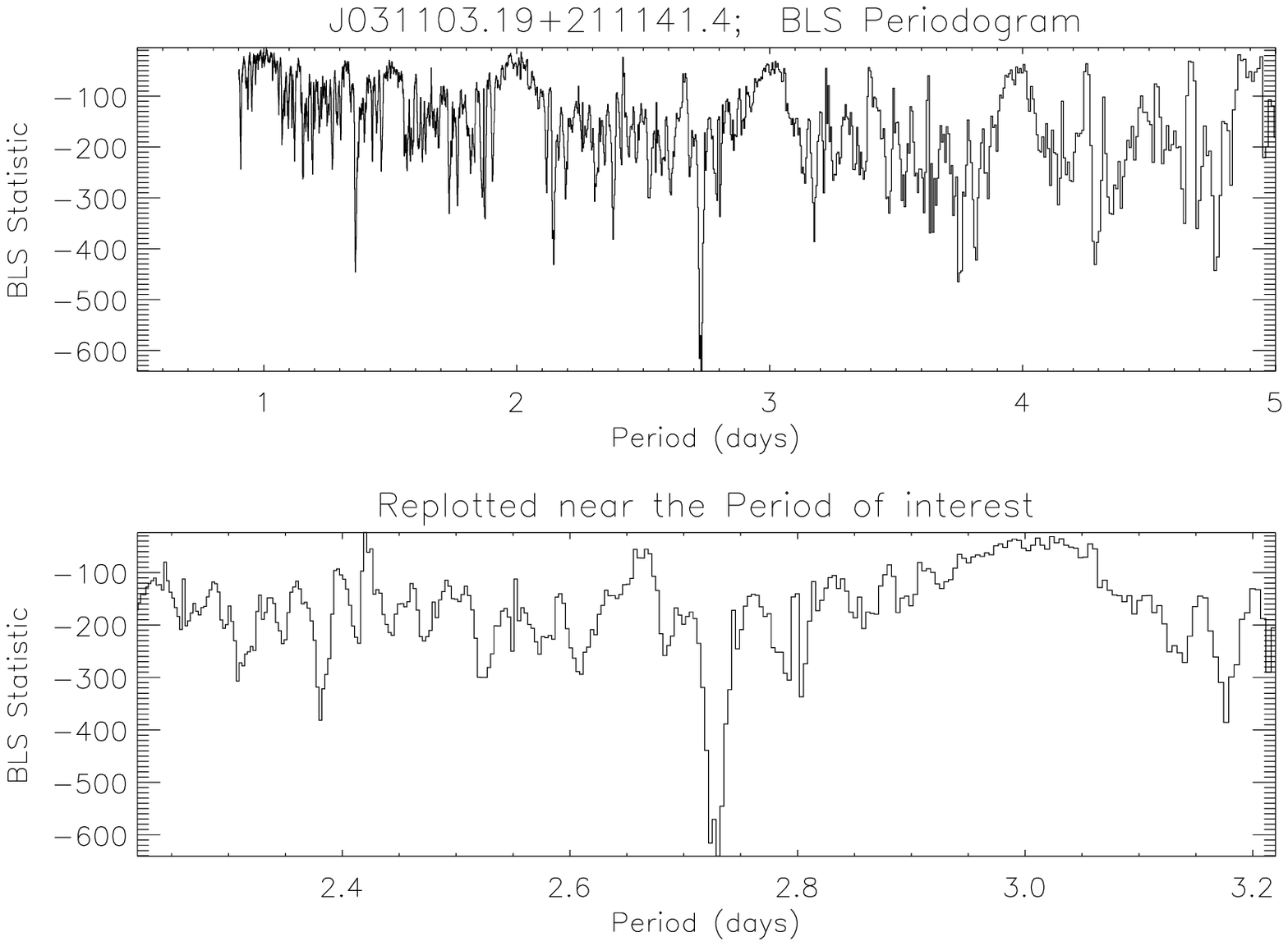,width=8cm}
  }}
\vspace{2mm}
  \centerline{\hbox{
  \psfig{file=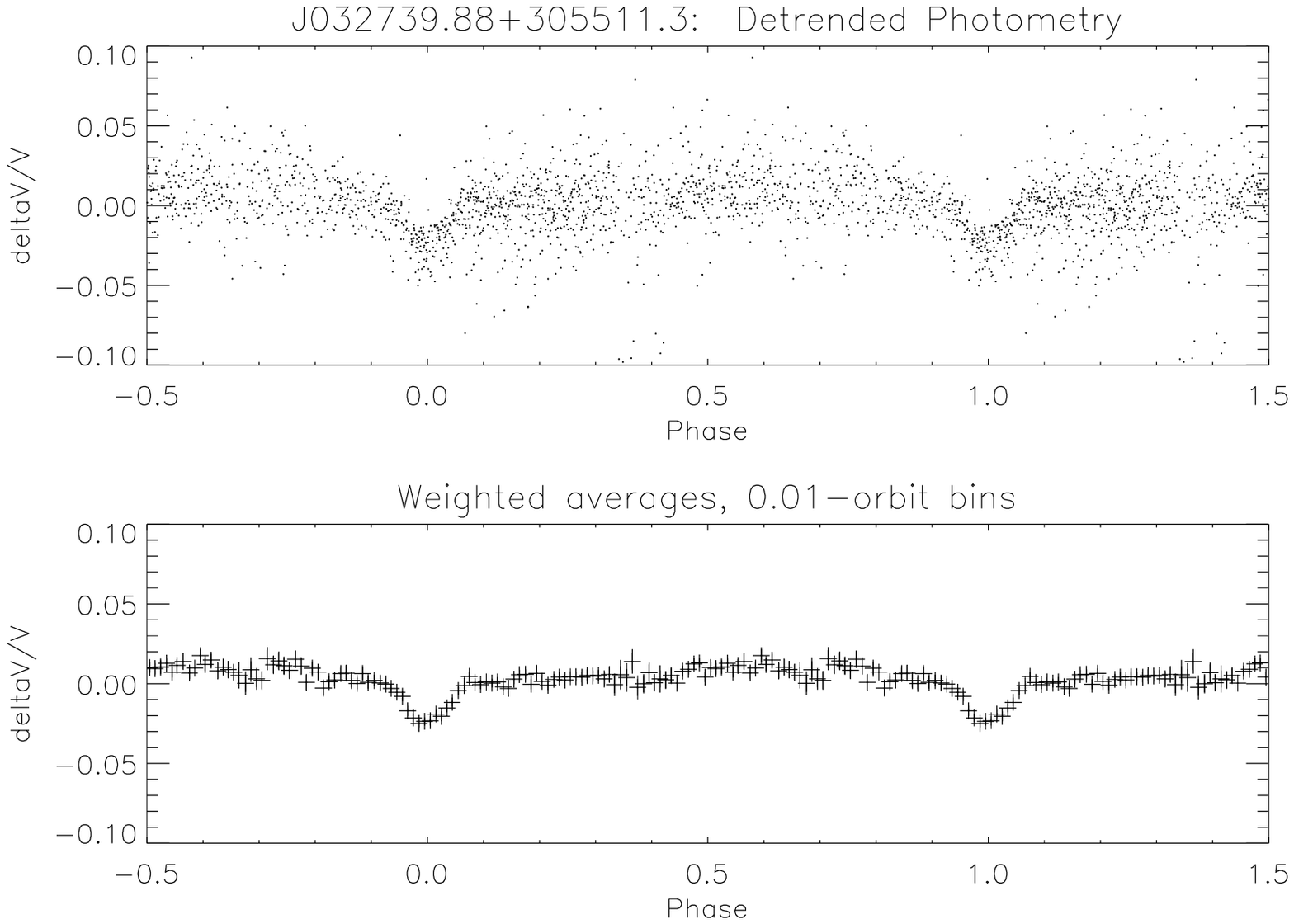,width=8cm}
  \psfig{file=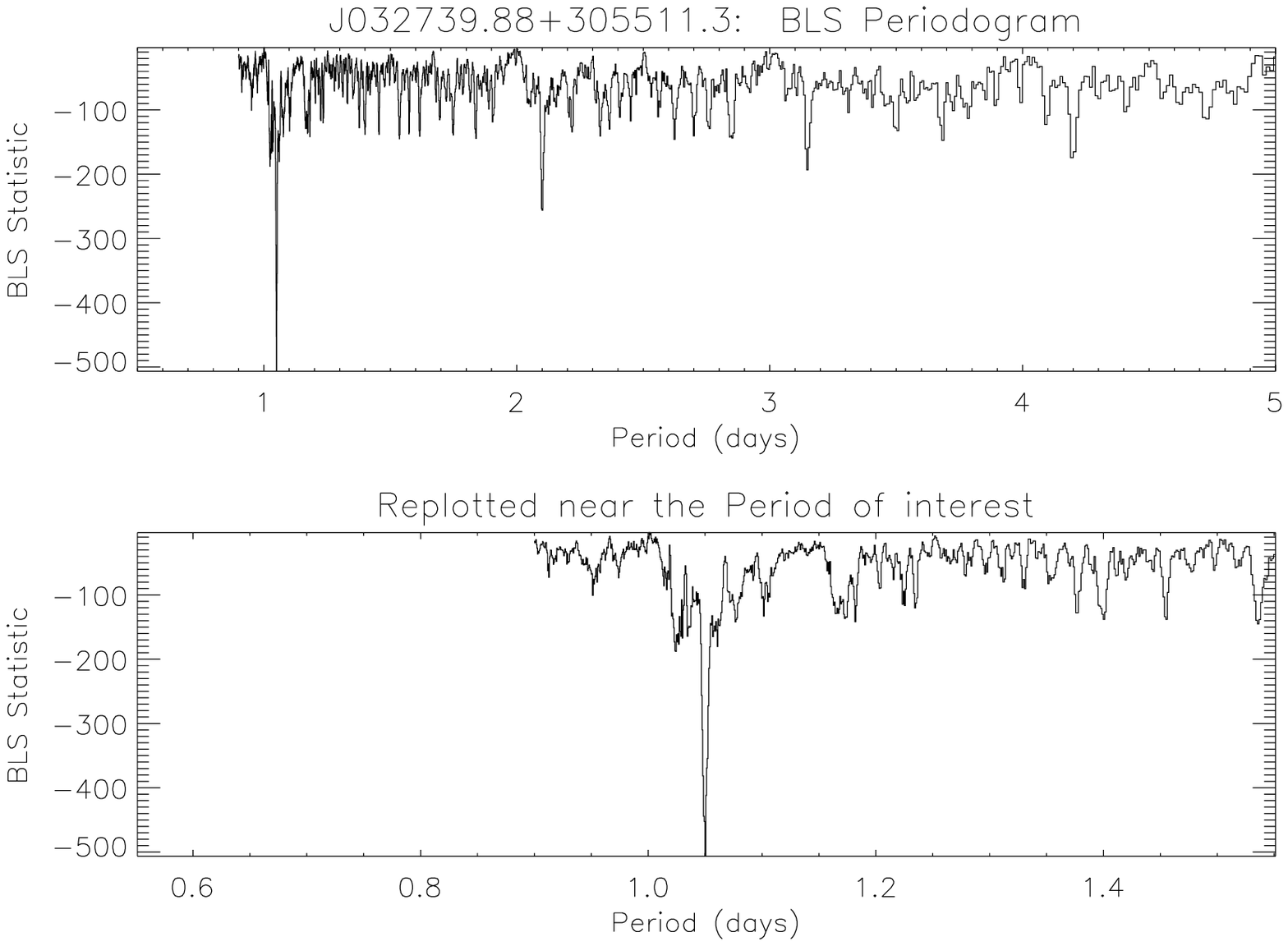,width=8cm}
  }}
\vspace{2mm}
  \centerline{\hbox{
  \psfig{file=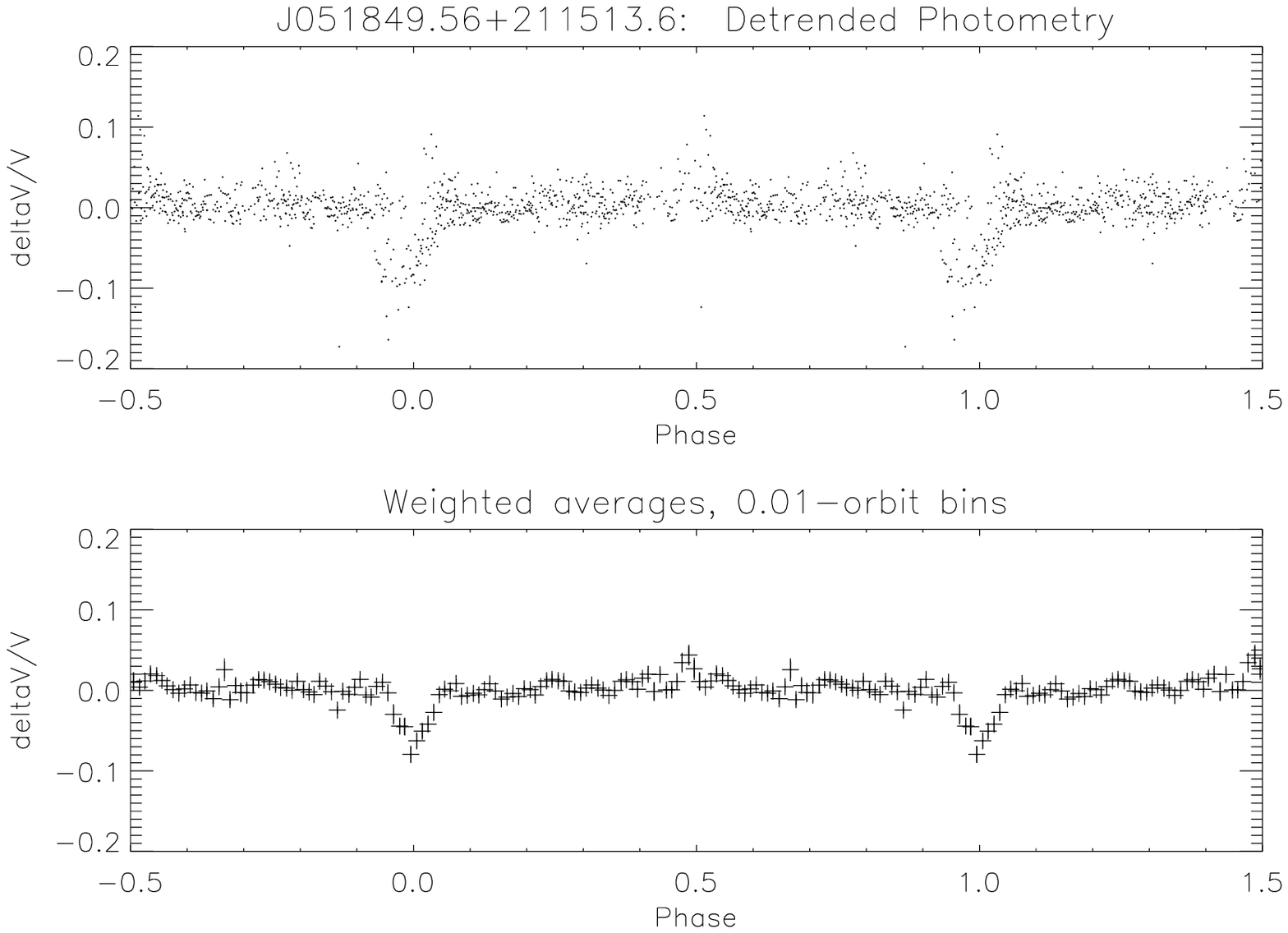,width=8cm}
  \psfig{file=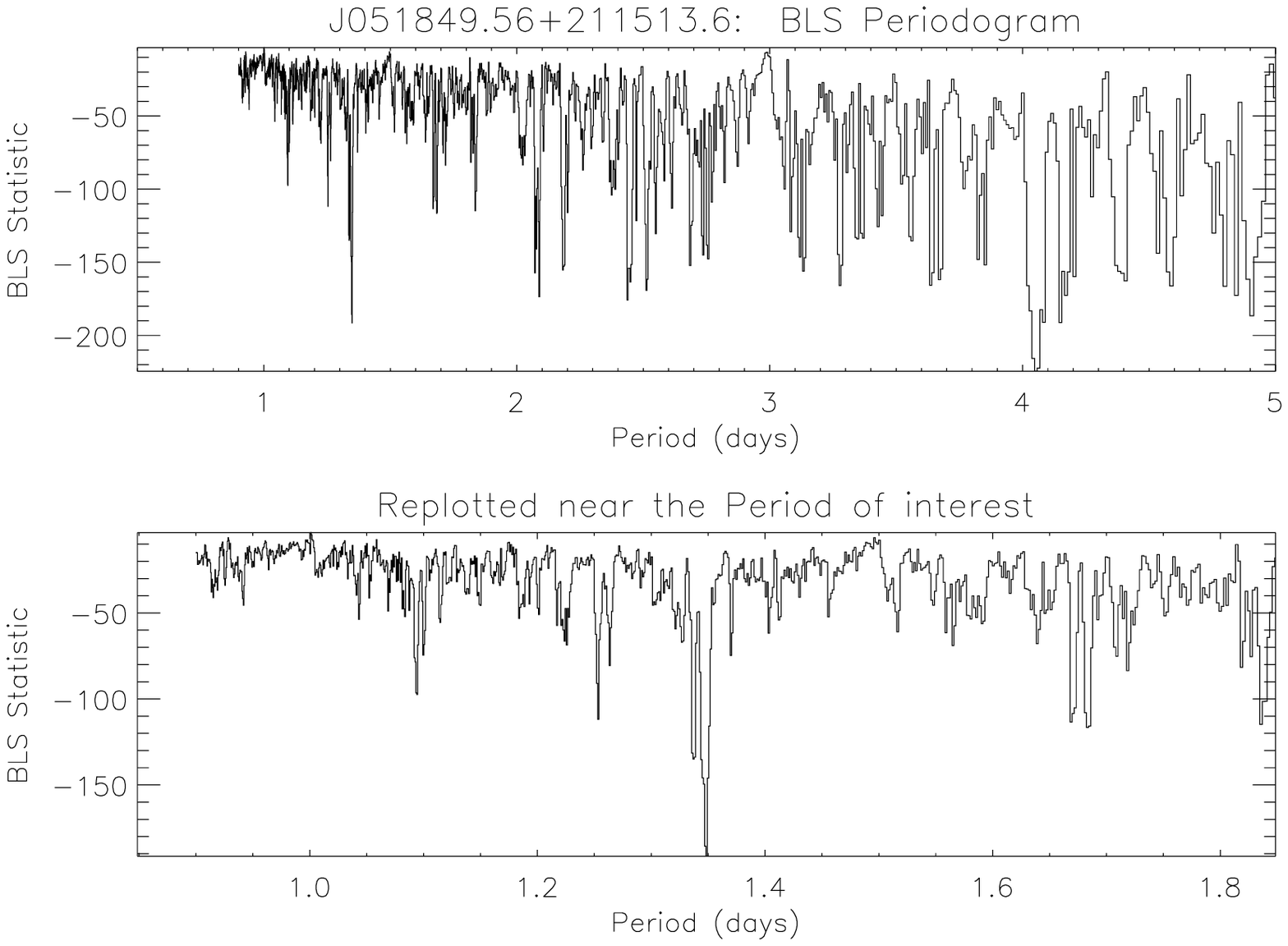,width=8cm}
  }}
%  \centerline{\hbox{
%      \psfig{file=J025922.67+275416.0_lc.ps,width=8cm}
%      \psfig{file=J025922.67+275416.0_pd.ps,width=8cm} 
%  }}
  \caption{The accepted Priority 2 Candidates: Folded lightcurves and BLS periodograms for (from top to bottom): J031103.19+211141.4; J032739.88+305511.3; J051849.56+211513.6.}
\label{pri2fig}
\end{figure*}

Finally we compute, but do not use as a selection criterion, the ratio
of observed transit width to that predicted given best-fit stellar
parameters, $\eta=W_{obs}/W$. In principle we expect genuine exoplanet
transits to show $\eta \sim 1$, with some range in values due to
observational scatter and inclination variations. This figure of merit
was introduced and computed for the OGLE transit candidates by Tingley
\& Sackett (2005); in practice all genuine OGLE transiting planets
show $0.5 \la \eta \la 1$.
%We estimate $H,K$ colours for each object by using
%the Carpenter (2001) transformations to express 2MASS $H,K_S$ in the
%Bessell \& Brett system (Bessell \& Brett 1988; see also Glass 1999
%and Carter 1990). The radius is inferred from the depth of transit
%$\delta$ and the stellar radius $R_*$ using the limb-darkened relation
%of Tingley \& Sackett (2005; their equation 9).

{\it Positional Matching:} We also visually check the positions
localised by the SW-N pipeline against catalogue position for the
target used by the automated query tool; in a few cases the measured
position was displaced by a small amount from the catalogue position
(even subpixel offsets can amount to nearly 15$''$; Section 2.1). Even
assuming perfect distortion-correction in the pipeline and no error
introduced in the conversion of positions between epochs in the
catalogues, objects with high proper-motion may have drifted
appreciably in the 2-3 decades since some of the catalogue
observations were made. In cases where an object is detected at a
slightly different location to its catalogue position, the automated
catalogue query tool can misidentify the target as a blending
neighbour. In these cases we use the measured magnitude $V_{SW}$ to
determine the most likely matching catalogue object, and re-calculate
the diagnostics accordingly.
%when assessing candidate
%suitability based on catalogued properties, we ensure the most likely
%matching object is used and the diagnostics re-calculated accordingly.
%Here the
%diagnostics were re-calculated for the most likely matching object. We
%provide specific examples of some of the more interesting
%false-positives below.  \\

% put the table here. 

{\it Crowding:} Candidates were rejected outright if any object
brighter than the candidate was present within the 48$''$ SW-N
aperture. For candidates with nearby objects {\it fainter} than the
candidate, we calculate the magnitude of the nearby object that would
be required for a 50\%-depth eclipse from the object to produce the
observed transit-depth from the aperture; if this magnitude is
surpassed the candidate is rejected.

\section{Results}

The bottom line of this analysis is that one out of 2688 candidates is
put forward as a Priority 1 target for spectroscopic follow-up with
three of 2688 Priority 2 targets. Table 1 gives the field statistics
for the search. Table 2 lists the 44 objects surviving visual
inspection, Table 3 gives the four candidates finally accepted. We
provide notes on the accepted objects below, as well as a subset of
the rejected candidates.
%Most of the rejected objects can be quickly
%dispensed with; reasons for rejection are given in tables 4 and 5 in
%the appendix. 
Some of the rejected objects are of interest in their own right,
either because their rejection is illustrative of the procedures we
followed to filter out candidates, or because the objects are
astrophysically interesting (Section 4.3).

\subsection{Priority 1 Candidate}

Only one object assigned Priority 1 on the basis of the visual and
S/N cuts (Section 3.2) survived the application of catalogue
information. See Figure \ref{pri1fig} for its lightcurve and BLS
periodogram.

%{\bf 1SWASP J031632.80+300144.2:} Shows six transit-like events of
%$\sim3.1\%$; at this signal-to-noise the folded profile is not
%entirely symmetric, though this may still be an artefact of the
%reduction. The 2.199-day period is clearly distinct from any aliases
%in the periodogram, and photometric colours imply late-G/early-K-type
%main-sequence parent star. 
% I THINK THIS OBJECT SHOWS ELLIPSOIDAL VARIATIONS!!

\begin{figure*}
  \centerline{\hbox{
 \psfig{file=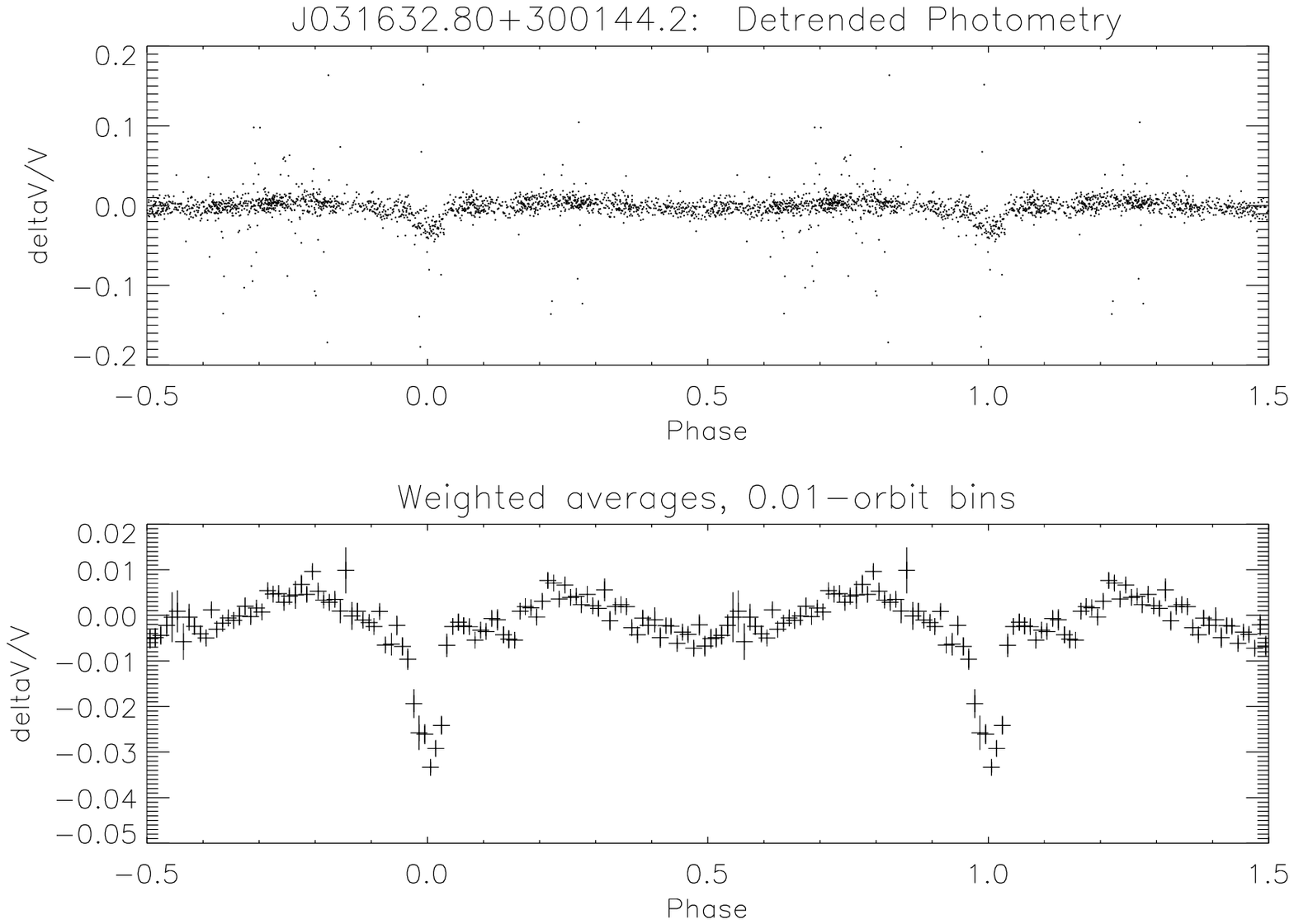,width=8cm}
 \psfig{file=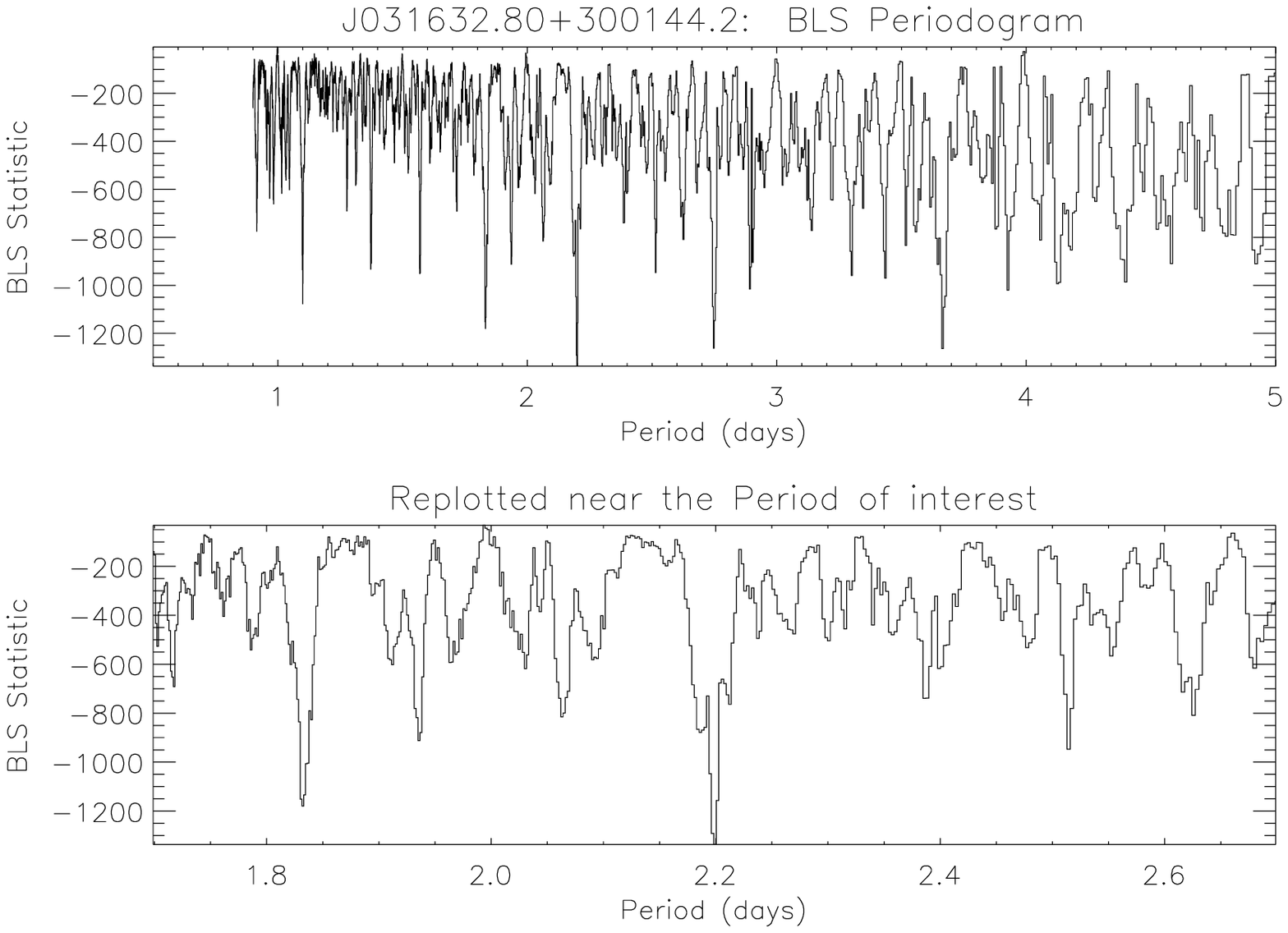,width=8cm}
}}
\caption{Phase-folded lightcurve and BLS periodogram of the otherwise
  excellent transit candidate 1SWASP J031632.80+300144.2. {\it Left:}
  the phase-folded lightcurve as it would appear at the visual
  examination stage (top) and re-plotted on a compressed flux-scale
  (bottom); the existence of ellipsoidal variations is clear. (Compare
  with Figure 1 of Sirko \& Paczy\'{n}ski 2003.) {\it Right:} the BLS
  periodogram.}
\label{ellip}
\end{figure*}

\begin{figure*}
 \centerline{\hbox{
      \psfig{file=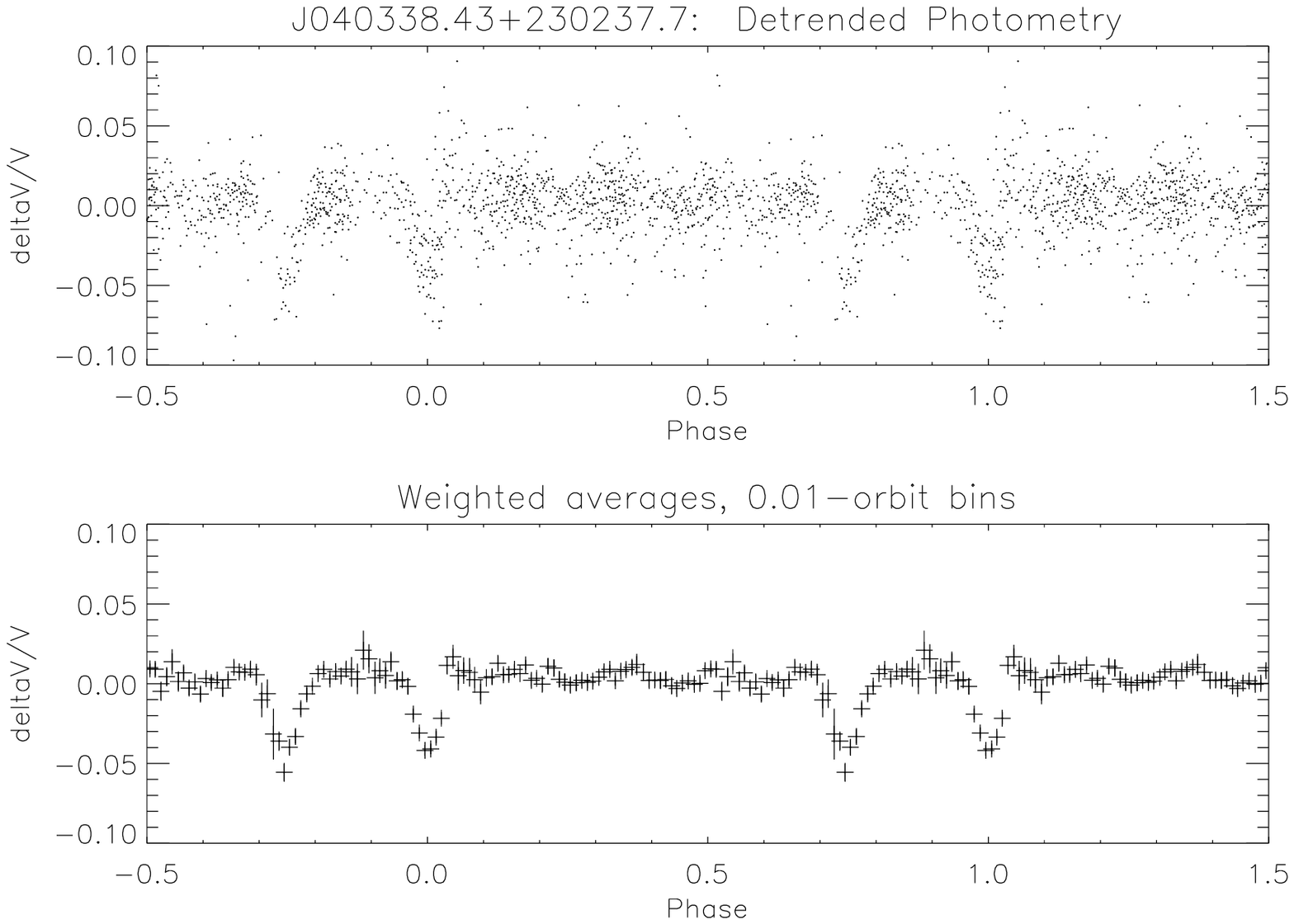,width=8cm} 
      \psfig{file=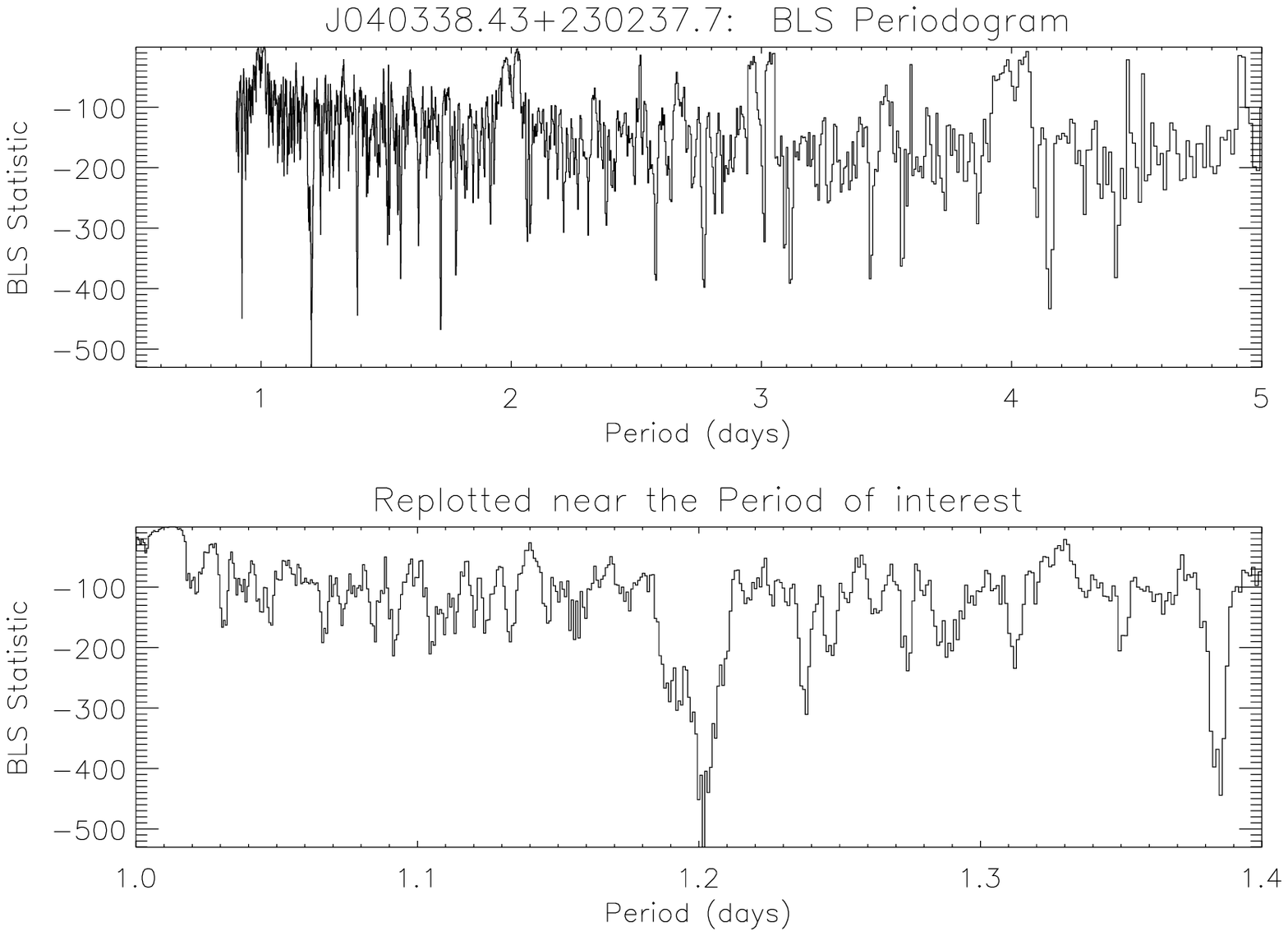,width=8cm} 
  }}
 \caption{Folded lightcurve and periodogram for the possible eccentric-orbit system J040338.43+230237.7.}
\label{eccentric}
\end{figure*}

{\bf 1SWASP J051221.34+300634.9:} 
%In initial investigation this object
%was originally set at Priority 2 due to a possibly V-shaped transit
%and the appearance of possible asymmetry in the lightcurve near
%transit; however on subsequent comparison with 
This object shows an almost prototypical transit-candidate event of
3$\%$-depth over a flat out-of-transit lightcurve. Five transits are
observed with a 1.24-day period and 1.87-hour transit duration. With
reduced proper motion 2.19 and ({\it V-$K_S$})= 1.61, this object is
firmly in the dwarf regime (Figure 2). Two USNO objects are $\ga$4.37
magnitudes fainter than the target within the 48$''$ SW-N aperture,
and thus are too faint to produce blending at the detected transit
level. 2MASS ($J$-$H$)=0.26 suggests a 1.15 $R_{\odot}$ F9 primary,
implying planet radius 1.71 $R_{J}$ and Tingley \& Sackett
$\eta$=0.77.

%Transformed ({\it H-K})=0.07 suggests a late-G-type
%main-sequence parent star, and the estimated stellar radius (0.98
%R$_{\odot}$) suggests a planetary radius 1.50 $R_J$.

\subsection{Priority 2 Candidates}

Two objects were initially assigned Priority 1 from visual analysis
and the S/N cuts in Section 3.2; however inclusion of prior
information from catalogues highlighted some uncertainty in the
luminosity class of these objects, thus they were demoted to
Priority 2. One further object that was initially assigned
Priority 2, survived the inclusion of catalogued
information. Periodograms and lightcurves for all three objects can be
found in Figure \ref{pri2fig}.

%\begin{figure*}
%  \centerline{\hbox{
% \psfig{file=J031621.80+300144.2_lc.ps,width=8cm}
% \psfig{file=J031621.80+300144.2_pd.ps,width=8cm}
%}}
%\caption{Phase-folded lightcurve and BLS periodogram of the otherwise
%  excellent transit candidate 1SWASP J031632.80+300144.2. {\it Left:}
%  the phase-folded lightcurve as it would appear at the visual
%  examination stage (top) and re-plotted on a compressed flux-scale
%  (bottom); the existence of ellipsoidal variations is clear. (Compare
%  with Figure 1 of Sirko \& Paczy\'{n}ski 2003.) {\it Right:} the BLS
%  periodogram.}
%\label{ellip}
%\end{figure*}
%
%
%\begin{figure*}
% \centerline{\hbox{
%%      \psfig{file=J053228.77+323620.3_lc.ps,width=8cm}
%      \psfig{file=J040338.43+230237.7_lc.ps,width=8cm} 
%      \psfig{file=J040338.43+230237.7_pd.ps,width=8cm} 
%  }}
% \caption{Folded lightcurve and periodogram for the possible eccentric-orbit system J040338.43+230237.7.}
%\label{eccentric}
%\end{figure*}

{\bf 1SWASP J031103.19+211141.4:} Deep (4.03\%), clearly visible
transit events are present, though there may be some structure in the
transit besides a planetary-type event. No Tycho-2 or {\it Hipparcos}
objects are found at the object position or within a 48$''$ radius,
making stellar radius determination using the apparent diameter
relations of Kervella et al (2004) impossible. USNO lists no potential
blends within the SW-N aperture. The luminosity class of this object
is somewhat open to question. USNO reports a bad measurement for
proper-motion, so the reduced proper motion has nothing to say about
the luminosity class of this object (at {\it V-$K_S$}=1.88 this
measure would be ambiguous for this object for proper-motions $\la
8$mas yr$^{-1}$). 

%transformed {\it (J-H),(H-$K_S$)}=0.32,0.033 respectively, near-IR colors
%are also inconclusive (Figure 5 of Bessell \& Brett 1988).

{\it Assuming} the parent star is luminosity-class $V$, 2MASS
($J$-$H$)=0.27 implies parent spectral type G0 and radius $\sim$1.12
$R_{\odot}$. This implies a planetary radius $\sim$1.89 $R_J$, with
Tingley \& Sackett figure of merit $\eta$=0.94, just within the range
corresponding to likely exoplanet transits (Tingley \& Sackett 2005).

%Using a H-K/J-H diagram, with H-K = 0.061 and
%J-H = 0.269, the star sits firmly on the Dwarf branch. The spectral
%type of the target is unclear; from V-K colour/temperature
%relationship, it could be G6-K0, and using J-H, it could be
%F5-F7. Taking a stellar radius of 0.9 R$_{\odot}$, this would imply a
%companion with radius 1.52 R$_j$, with an $$\eta$$ of 1.21. 

\begin{figure*}
 \centerline{\hbox{
  \psfig{file=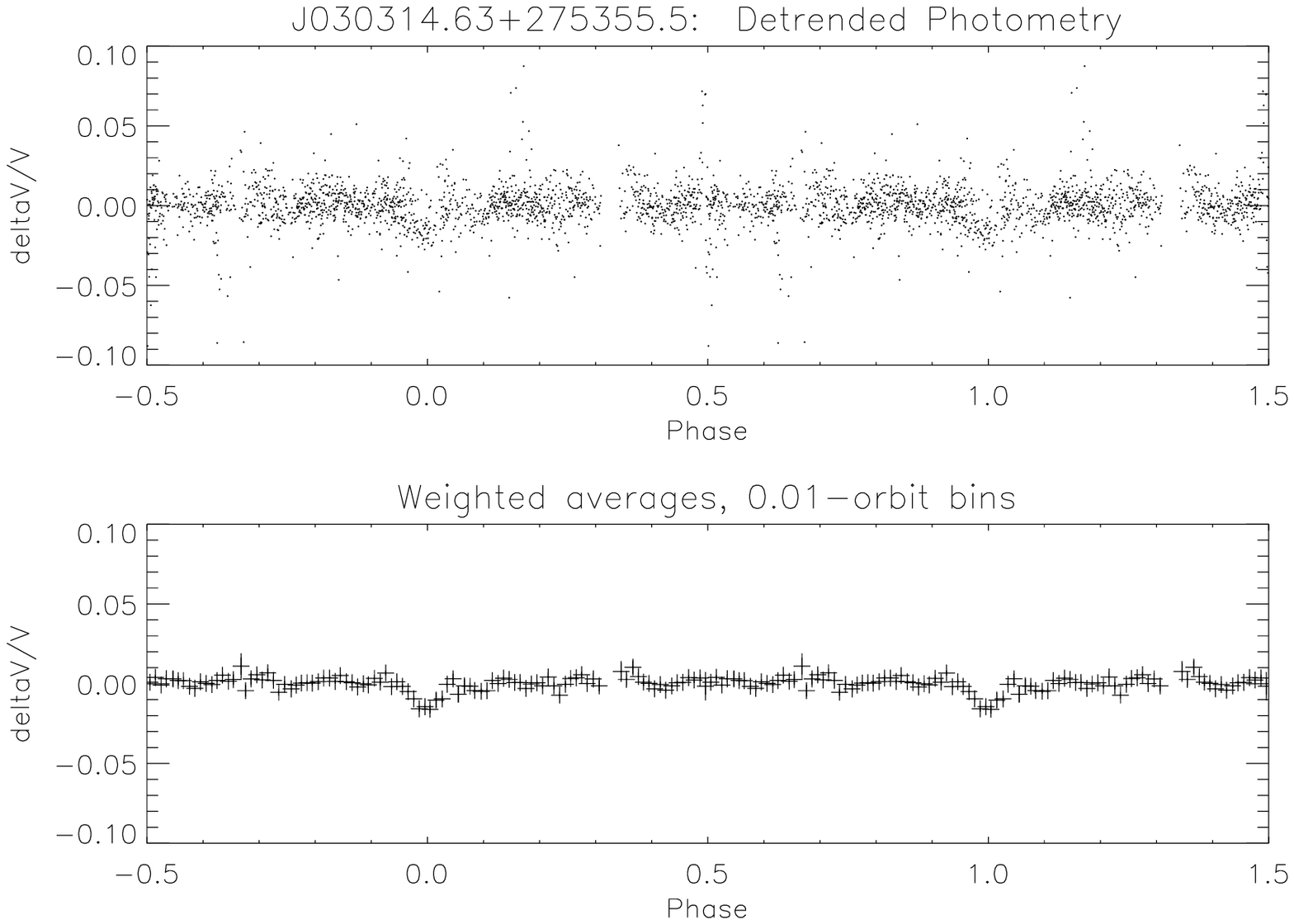,width=8cm}
  \psfig{file=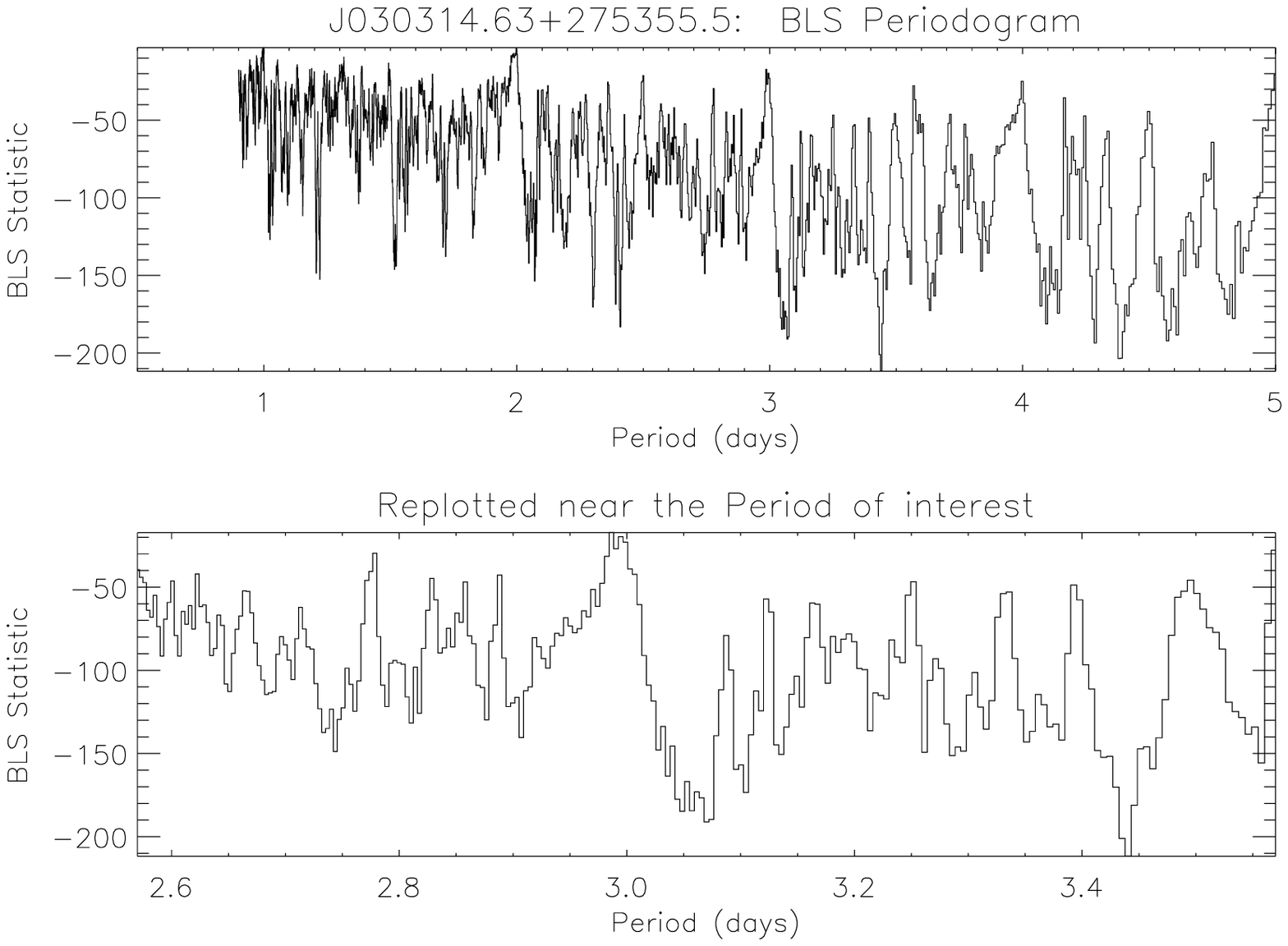,width=8cm}
  }}
 \centerline{\hbox{
  \psfig{file=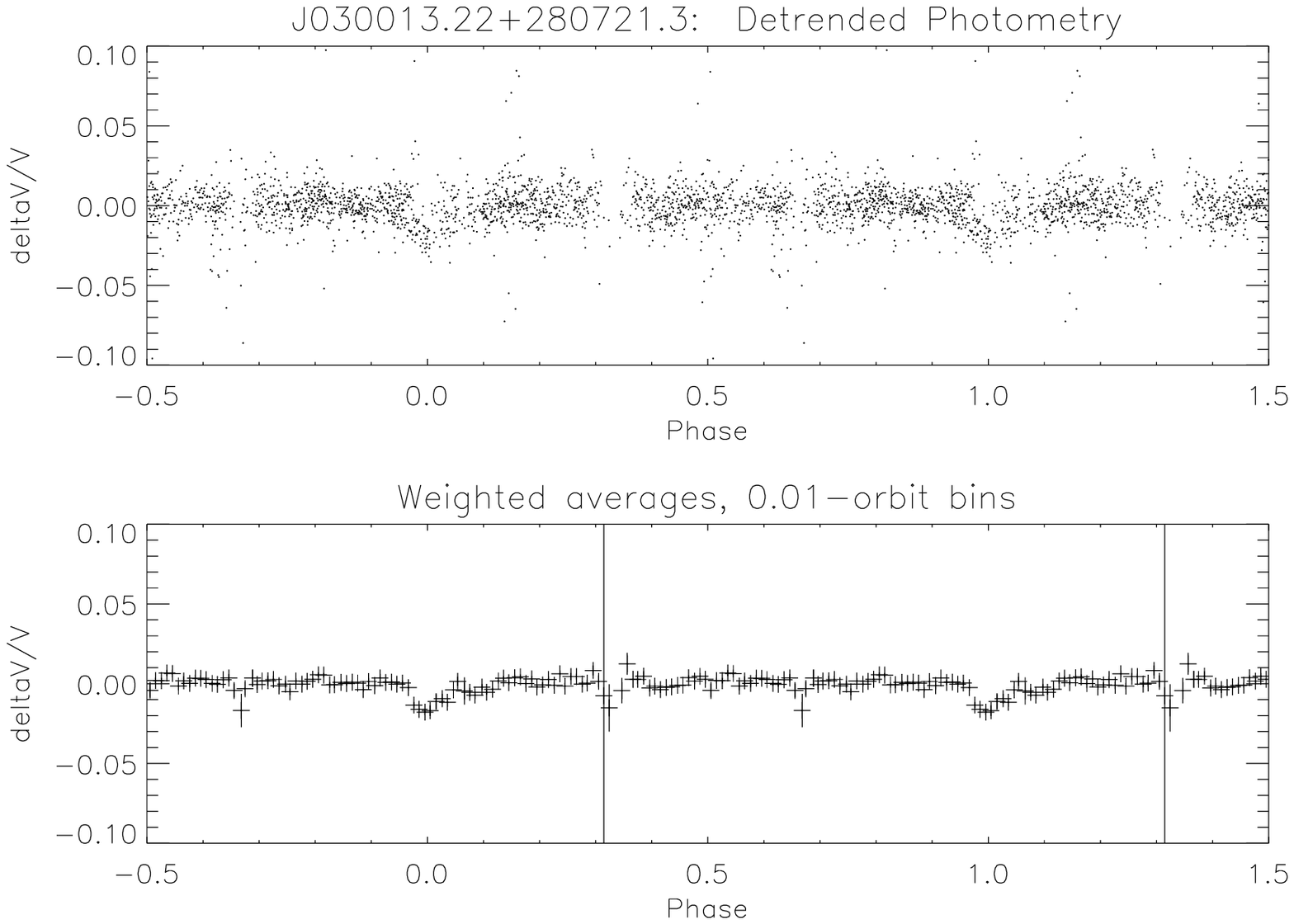,width=8cm}
  \psfig{file=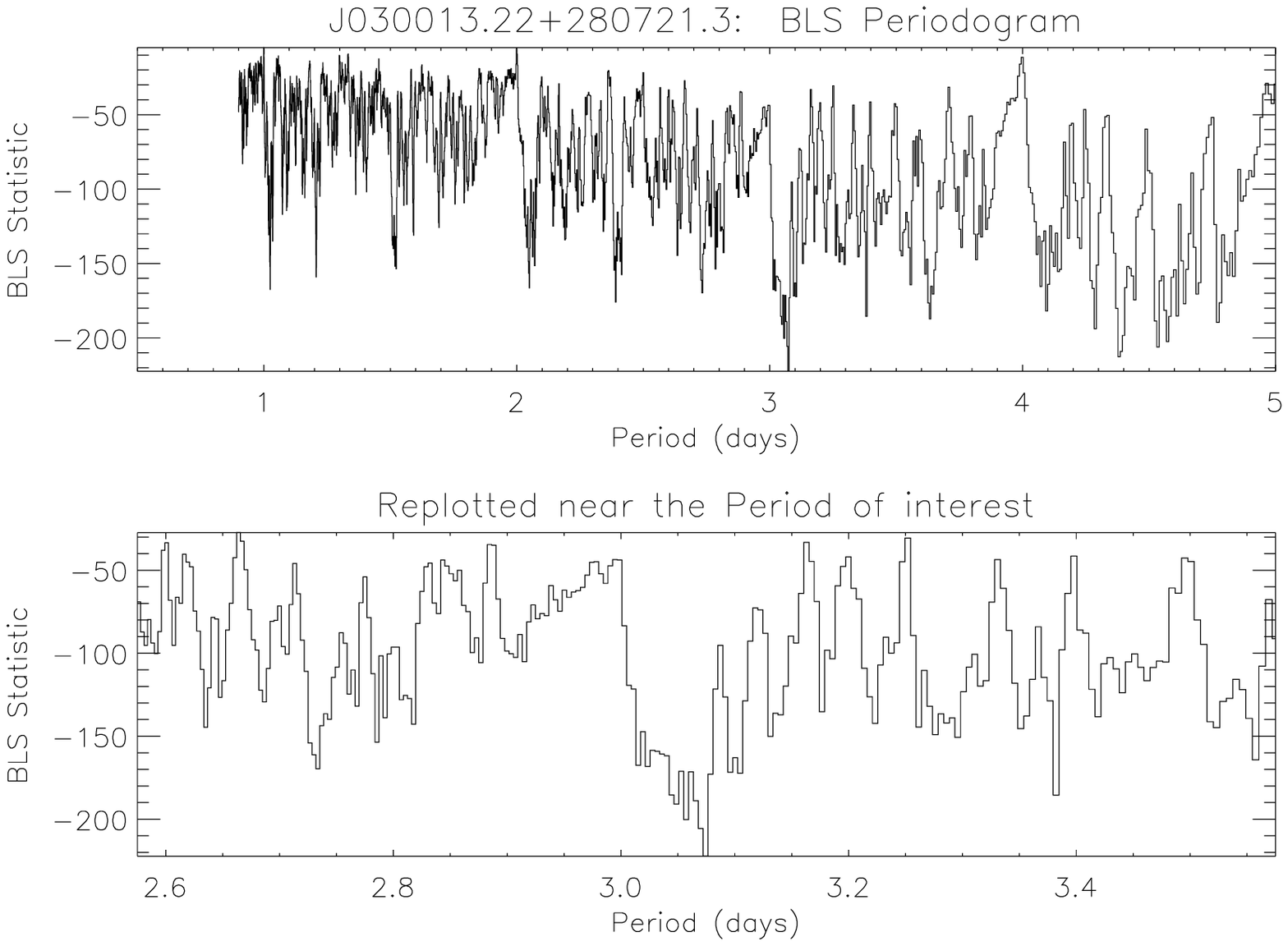,width=8cm}
  }}
 \centerline{\hbox{
  \psfig{file=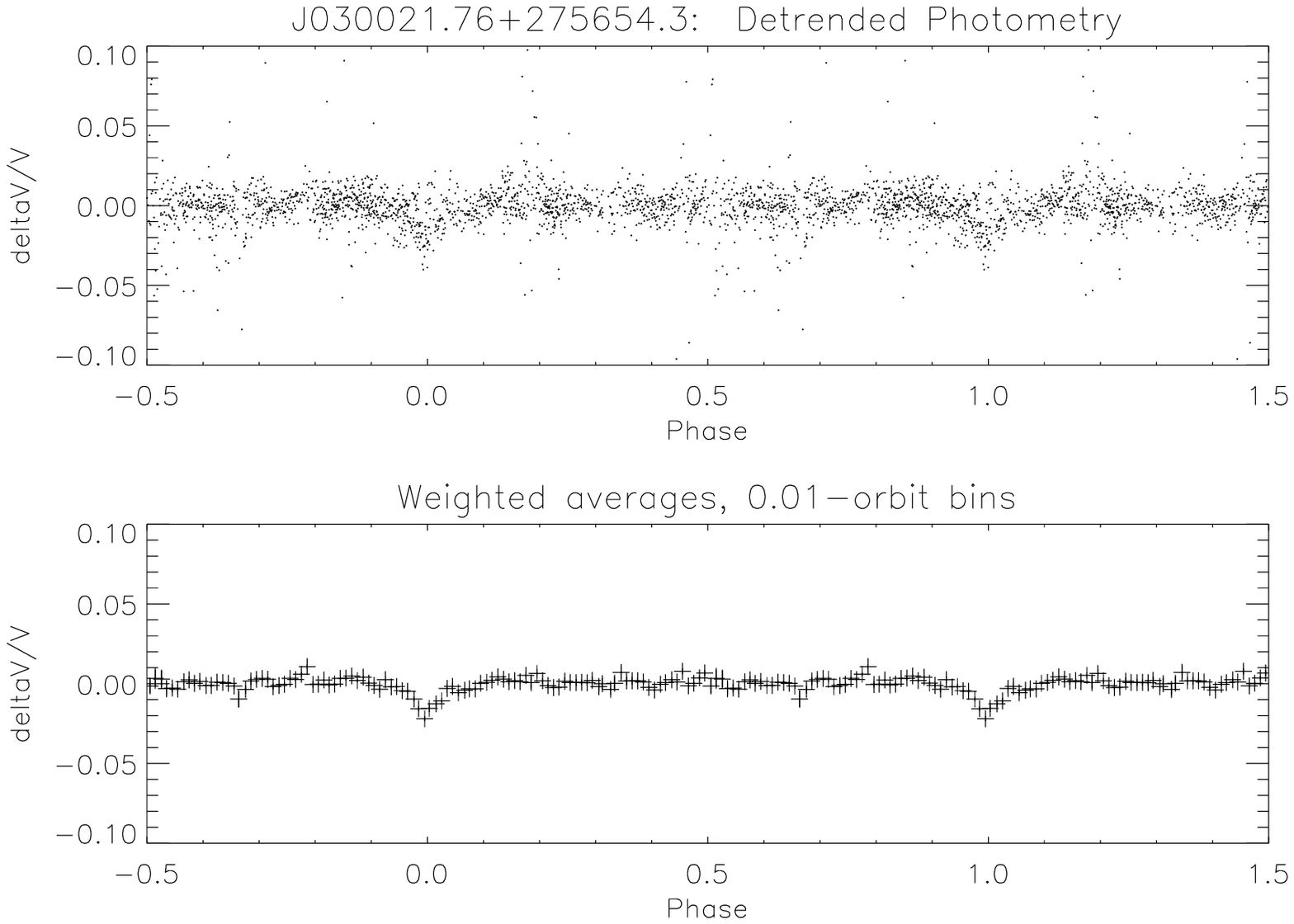,width=8cm}
  \psfig{file=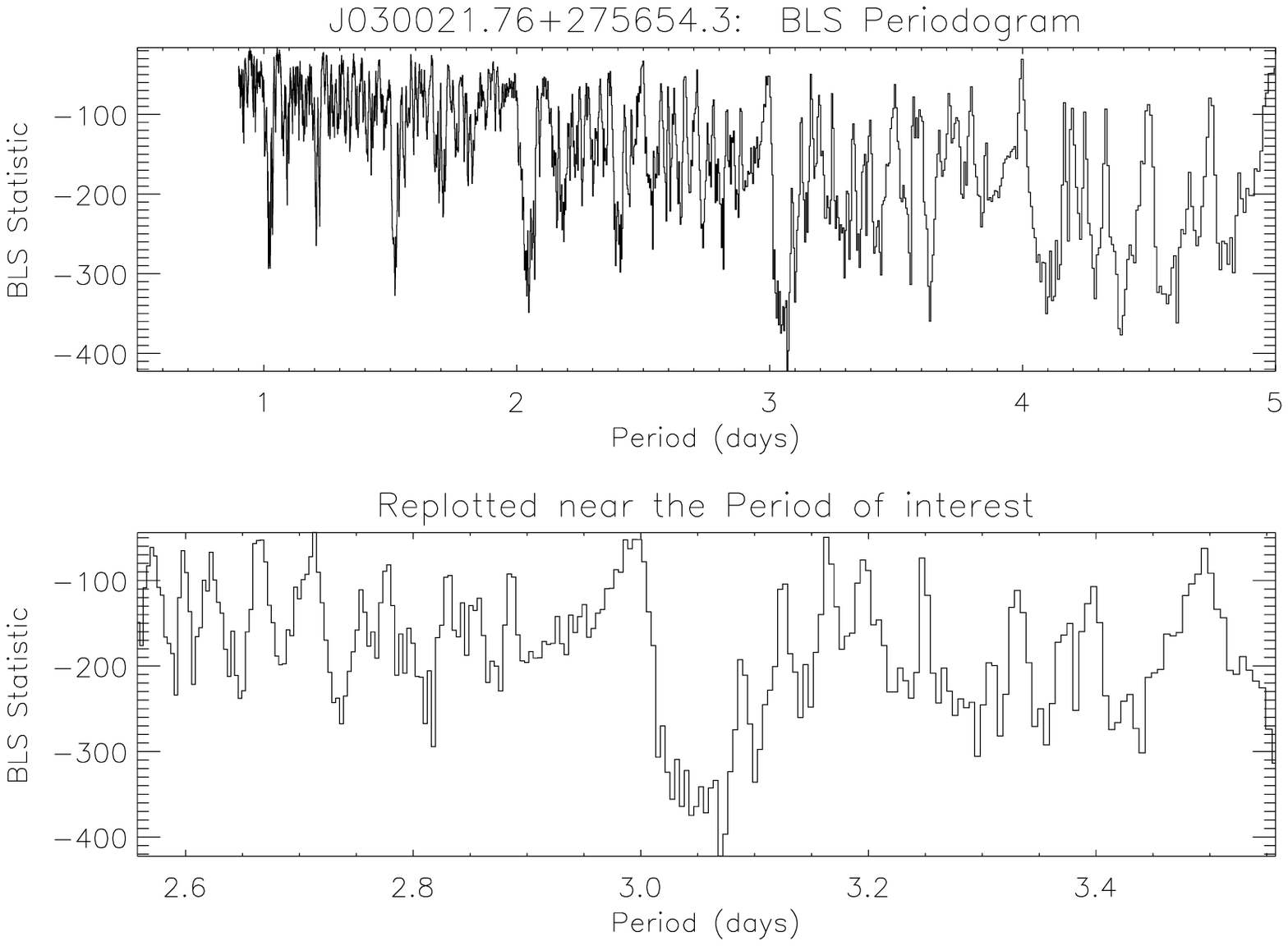,width=8cm}
  }}
\caption{Folded lightcurves and periodograms (after detrending) for
  the objects J030314.63+275355.5 (top), J030013.22+280721.3 (middle)
  and J030021.76+275654.3 (bottom). Detected periods are 3.07125,
  3.070201 and 3.0577 days respectively (from top to bottom). The
  nearest pair are 10.7 arcmin (43 pixels) from each other but show such
  similar folded lightcurves and periods that we reject all three as
  possible candidates.}
\label{artefact}
\end{figure*}

{\bf 1SWASP J032739.88+305511.3:} This object shows a clear shallow
transit-like event (2.53\%), on a 1.05 day period clearly distinct
from the 1-day trough in the periodogram (Figure \ref{pri2fig}); at
this period, fourteen transits are observed. No {\it Hipparcos} or
Tycho-2 objects are found at the target position or within 48$''$ of
the target, so direct inference of the stellar radius (c.f. Kervella
et al. 2004) is not possible. USNO lists no potential blending objects
within the SW-N aperture. As with J031103.19+211141.4, the USNO proper
motion measurement cannot be used due to poor quality (Section
3.3). However, with ({\it V-$K_S$})$\sim$2.9, this object would have
to show proper-motion $\ga$15 mas yr$^{-1}$ to be close enough to be a
likely dwarf (Figure 1), which is rather high to go unnoticed over the
25-year timebase of the USNO catalogue. Thus there is the suspicion
that this object may be a giant and it was thus demoted to Priority
2. Color index 2MASS ($J$-$H$)=0.22 suggests a 1.23 $R_{\odot}$ parent
with spectral type F8, {\it assuming} it falls on the
main-sequence. This predicts planet radius 1.69 $R_J$ and Tingley \&
Sackett $\eta$=1.06.

% switch figures again! 
%\begin{figure}
%\psfig{file=artefacts_v2.ps,width=8cm}
%\caption{The spatial distribution of objects in field SW3016+3126 with periods $3.06 \pm 0.015$ days (of which three example lightcurves are plotted in Figure \ref{artefact}), overlaid on a Digitized Sky Survey image (North to top, East to left; the figure measures $90' \times 60'$ and the objects are marked with circles). No obvious pattern (such as proximity to a bright source) is seen in the spatial distribution of artefacts. (One further object $\sim 60'$ to the East of this image is not shown).}
%\label{locations}
%\end{figure}

{\bf 1SWASP J051849.56+211513.6: } This object shows six transit-like
events of $\sim6\%$ depth on a 1.35-day period and with a 2.3-hour
transit duration. The transit lightcurve is rather deep and possibly
V-shaped, however consistent with a planetary transit given the
photometric precision (Figure \ref{pri2fig}). With reduced proper
motion $\sim 1.62$ and ({\it V-$K_S$})=2.2, this object lies within a
region of parameter space roughly equally populated by dwarfs and
giants, thus its luminosity class is uncertain. {\it Assuming} the
parent star is a main-sequence object, the 2MASS colors ($J$-$H$)=0.27
suggest a 1.12 $R_{\odot}$ G0 primary, implying planetary radius 2.3
$R_{J}$ and Tingley \& Sackett $\eta$=0.88, so these parameters are
consistent with a transiting exoplanet. In addition to the
luminosity-class uncertainty for this object, the lightcurve shows
possible variability at anti-transit, and when folded on the most
significant BLS period-detection (4.05 days) only shows two transits
(we used the next most-significant period of 1.35 days in this
analysis). This object is thus kept at Priority 2 pending further
lightcurve sampling in the upcoming 2006 dataset.

\subsection{Example Rejected Candidates}

% now going through again and re-checking all the descriptions!

{\bf 1SWASP J031632.80+300144.2 - ellipsoidal variations:} This object
shows six transit-like events of $\sim3.1\%$ depth; at this
signal-to-noise the folded profile (Figure \ref{ellip}) is not
entirely symmetric in the region of the transit, though this may still
be an artefact of the reduction. The 2.199-day period is clearly
distinct from any aliases in the periodogram, and photometric colours
imply late-G/early-K-type main-sequence parent star. However, the
signal-to-noise of ellipsoidal variations is high, at $\sim$26, and
indeed a re-plotting of the lightcurve on a wider phase-scale and
compressed flux-scale shows quite clearly the existence of apparently
ellipsoidal variations (compare with e.g. Figure 1 of Sirko \&
Paczy\'{n}ski 2003). This object is probably a grazing-incidence
stellar binary.

% I THINK THIS OBJECT SHOWS ELLIPSOIDAL VARIATIONS!!

%\begin{figure}
% \psfig{file=depths.ps,width=8cm,angle=-90}
%\caption{Transit depths of the five candidates. Asterisks - priority one, squares - priority two.}
%\end{figure}

%\begin{figure}
%\psfig{file=artefacts.ps,width=8cm}
%\caption{The spatial distribution of objects in field SW3016+4126 with periods $3.06 \pm 0.015$ days (of which three example lightcurves are plotted in Figure \ref{artefact}), overlaid on a Digitized Sky Survey image (North to top, East to left; the dashed-line box is $90' \times 60'$ and the objects are marked with circles). No obvious pattern (such as proximity to a bright source) is seen in the spatial distribution of artefacts. (One further object $\sim 60'$ to the East of this image is not shown).}
%\label{locations}
%\end{figure}

{\bf J040338.43+230237.7 - eccentric-orbit binary?} This object shows
pairs of occultations at different depths when folded on the detected
period of $1.20 \pm ( 4.1 \times 10^{-5}$) days, but for which the
secondary events are far from phase 0.5 (Figure \ref{eccentric}). The
transit-like events are well-sampled, with seven transit-like events
observed. Further analysis of this interesting object will be reported
in a future paper.

{\bf J044639.17+394837.6 - blend:} This object is apparently very
heavily blended and its catalogue magnitudes (e.g. $B\sim16.5$; no
Tycho-2 $V$-magnitude is present for this object) are far from those
measured ($V_{sw}\sim12.0$). The nearby (22$''$) object USNO
1298-0108374 has Tycho-2 $V$ magnitude $V_{Ty2}\sim12.3 \pm 0.4$, much
closer to $V_{sw}$. It is surrounded by 8 objects within 3.5-5
magnitudes, however, which could contribute up to $\sim20\%$ of the
light in the aperture. Thus both candidate counterparts are too
blended to allow a planetary companion for the observed depth of
eclipse-like event.

% This one IS still in! 
{\bf J045349.66+333842.5 - X-ray faint H$\alpha$ emission-line
  object:} Rejected because two blended objects are within 5
magnitudes, {\it Simbad} shows this to be an $H\alpha$ emission-line
object. No $ROSAT$ source is detected. With a 1.8d period, this might
be an X-ray faint active binary with low-amplitude optical variability.

{\bf J040322.73+274841.5 - spectral type uncertainty, blend:} With
transit depth 2$\%$, this object exhibits 4 transits in the SW-N 2004
dataset against otherwise smooth behaviour outside ``transit.''  In
addition to the candidate, two red objects are found within the
aperture that are only two magnitudes fainter in $JHK_S$ and with USNO magnitude difference from the candidate $\Delta R \sim5.2,
\Delta I \sim3.5$. There is a third neighbour $<10''$ distant, but it
is roughly $\sim$7 mags fainter (by comparison with a well-separated
nearby object of similar apparent brightness in the DSS image). While
$(B-V) \sim0.3$ suggests spectral type roughly $\sim$F0 or so (Zombeck
1992), the $(V_{sw}-K)$ and $(J-H)_{2MASS}$ suggest spectral type
closer to late-F or early-G. As the SW-N bandpass includes Johnson
$RI$, (c.f. Figure 2 of Kane et al 2004), this object may well be
blended at the $1\%$ level in the SW-N bandpass.

\begin{figure}
\psfig{file=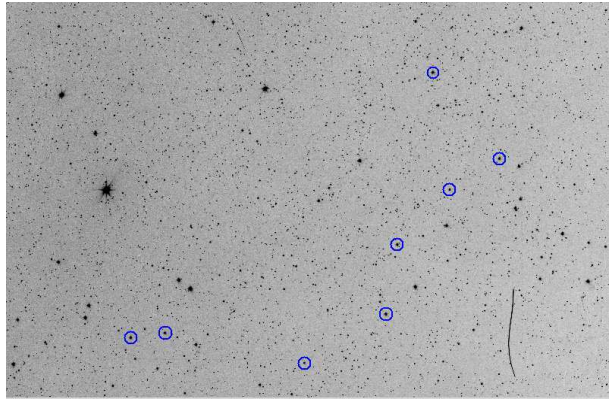,width=8cm}
\caption{The spatial distribution of objects in field SW3016+3126 with periods $3.06 \pm 0.015$ days (of which three example lightcurves are plotted in Figure \ref{artefact}), overlaid on a Digitized Sky Survey image (North to top, East to left; the figure measures $90' \times 60'$ and the objects are marked with circles). No obvious pattern (such as proximity to a bright source) is seen in the spatial distribution of artefacts. (One further object $\sim 60'$ to the East of this image is not shown).}
\label{locations}
\end{figure}

{\bf J030021.76+275654.3 - artefact:} This object was originally a
Priority 2 transit candidate. However its period and periodogram are
highly similar to a number of other distinct stellar objects in the
image. The zeropoints in the fitted ephemerides for each object are
highly similar - with an MJD$_0$ spread of only 1.5 hours - suggesting
the apparent period detection may have been dominated by lightcurve
artefacts still present after detrending (Figure \ref{artefact}). The
detrending and BLS period-search (section 3.2) produces a few hundred
exoplanet-candidates per field, which allows a simple check for
candidates that share the same period as several other objects, such
as J030021.71+27654.3. In principle the lightcurves of all candidates
might be examined with reference to the raw image, to determine if
such groups of candidates cluster near any bright, variable object or
along artefacts such as CCD bleeds. This requires a detailed, highly
accurate knowledge of the wings of the PSF both as a function of
on-chip position, and of frame-number. A simple plot of the position
of the candidates with similar periods shows that in fact these
objects are {\it not} near any extremely bright object; Figure
\ref{locations}.

\begin{figure*}
 \psfig{file=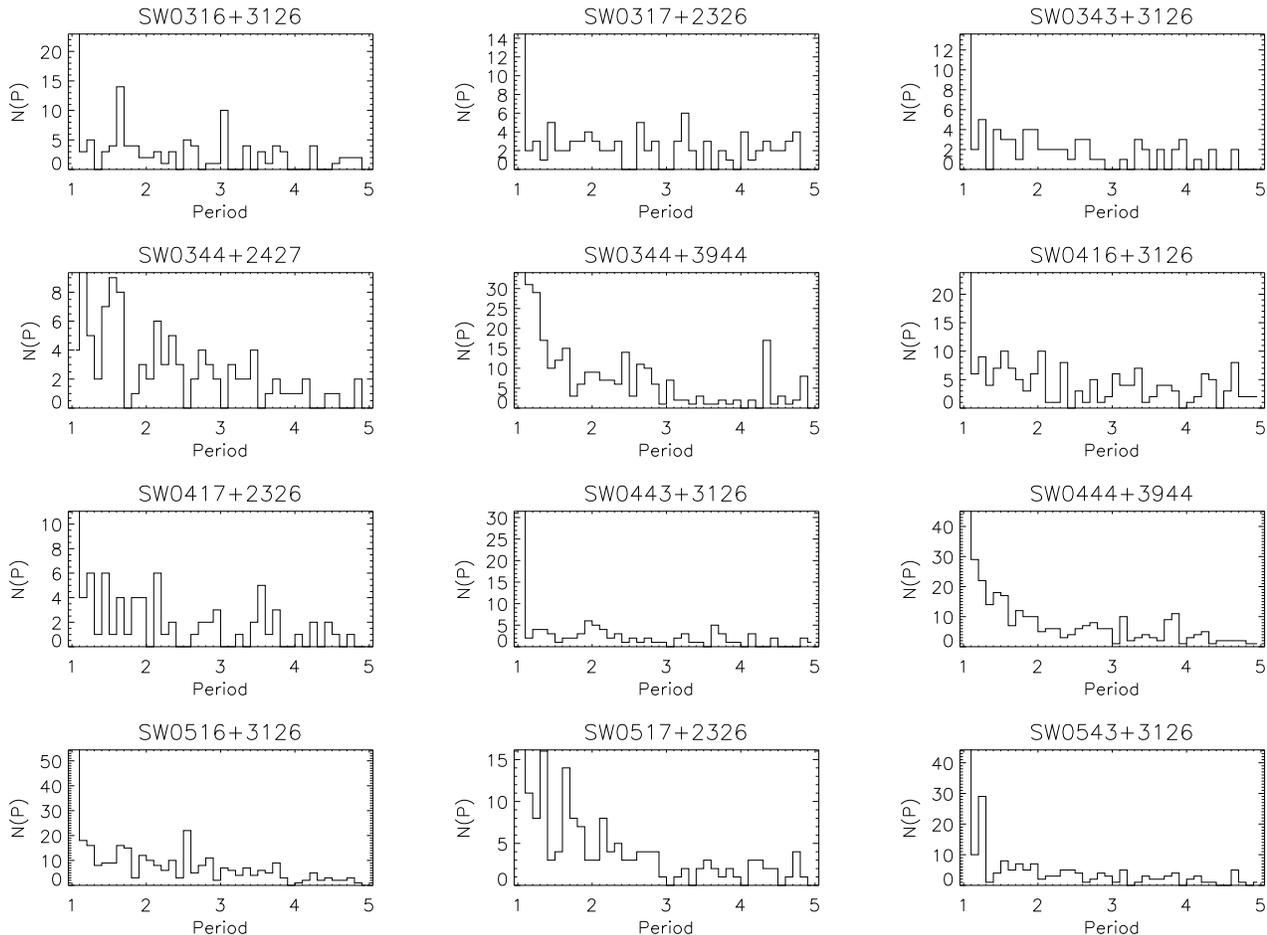}
 \caption{Periodicities returned from BLS period-search on lightcurves
   that have been detrended (Tamuz et al. 2005). Local increases in detected period represent possible shared-variability artefacts such as J030021.76+275654.3. The periods at which these likely artefacts occur are highly field-dependent. For example, the field SW0316+3126 (top left) shows ten objects with periods $3.06 \pm 0.015$ days. Examination of the lightcurves and periodograms of these objects shows a population of objects with shared variability, which must be removed from further consideration (Figure \ref{artefact}).}
\label{aliases}
\end{figure*}

A more efficient if cruder method is to search for peaks in the
distribution of detected periods to catch groups of highly similar
detected periods. The distribution of detected periods is highly
field-dependent (Figure \ref{aliases}), so we cannot sum over fields
to improve the statistics for this process. We see that for the field
containing J030021.76+275654.3 (field SW0316+3126) there is indeed a
rather high number (ten) of objects with detected period $3.06 \pm
0.015$ days. This highlights a number of other suspicious periods from
the fields observed. Candidates at these periods were {\it not}
rejected outright based on the period alone, but their lightcurves and
periodograms were compared to other objects with similar detected
period to screen for possible artefacts. The sample in Figure
\ref{aliases} contains lightcurves both with and without significant
sampling gaps - these are the lightcurves for all objects passed
forward for visual selection by the initial BLS period-search and
statistical cuts (Section 3.2). Objects where the automated search has
fit sampling-gaps or any residual nightly trends, cluster at 1-day
periods; these objects are not considered further. The general shape
of the distribution of detected periods does appear to broadly follow
the number of nights in each field, though the distribution of periods
detected clearly does not depend just on the number of nights alone (Figure
\ref{aliases}). For example, fields SW0344+2427 and SW0543+3126 show
similar drops to zero detections at or near periods of integer days,
and both are the most sparsely-sampled fields in this RA range (Table
\ref{stats}).

\begin{figure*}
 \centerline{\hbox{
 \psfig{file=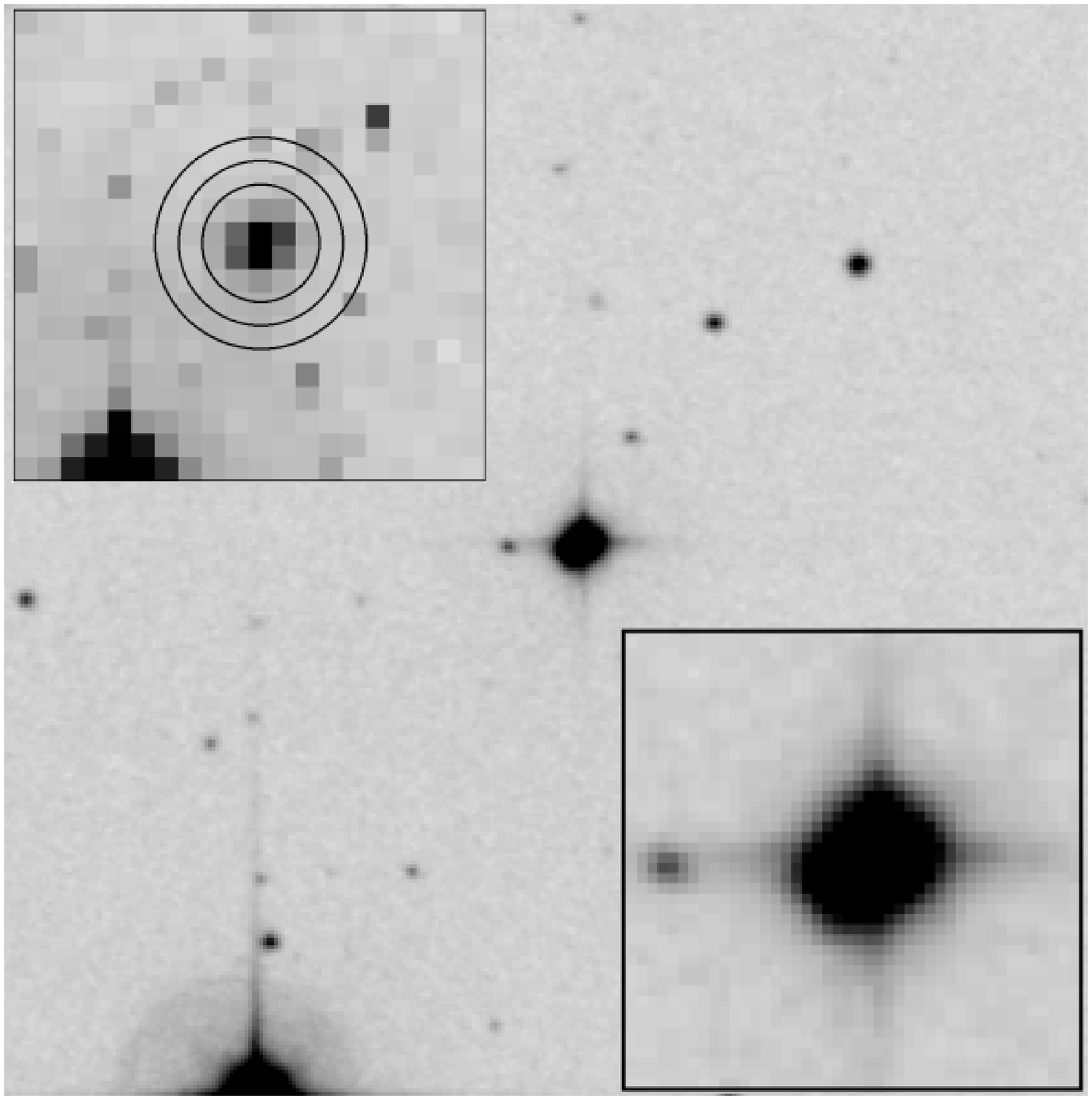,width=8cm}
 \psfig{file=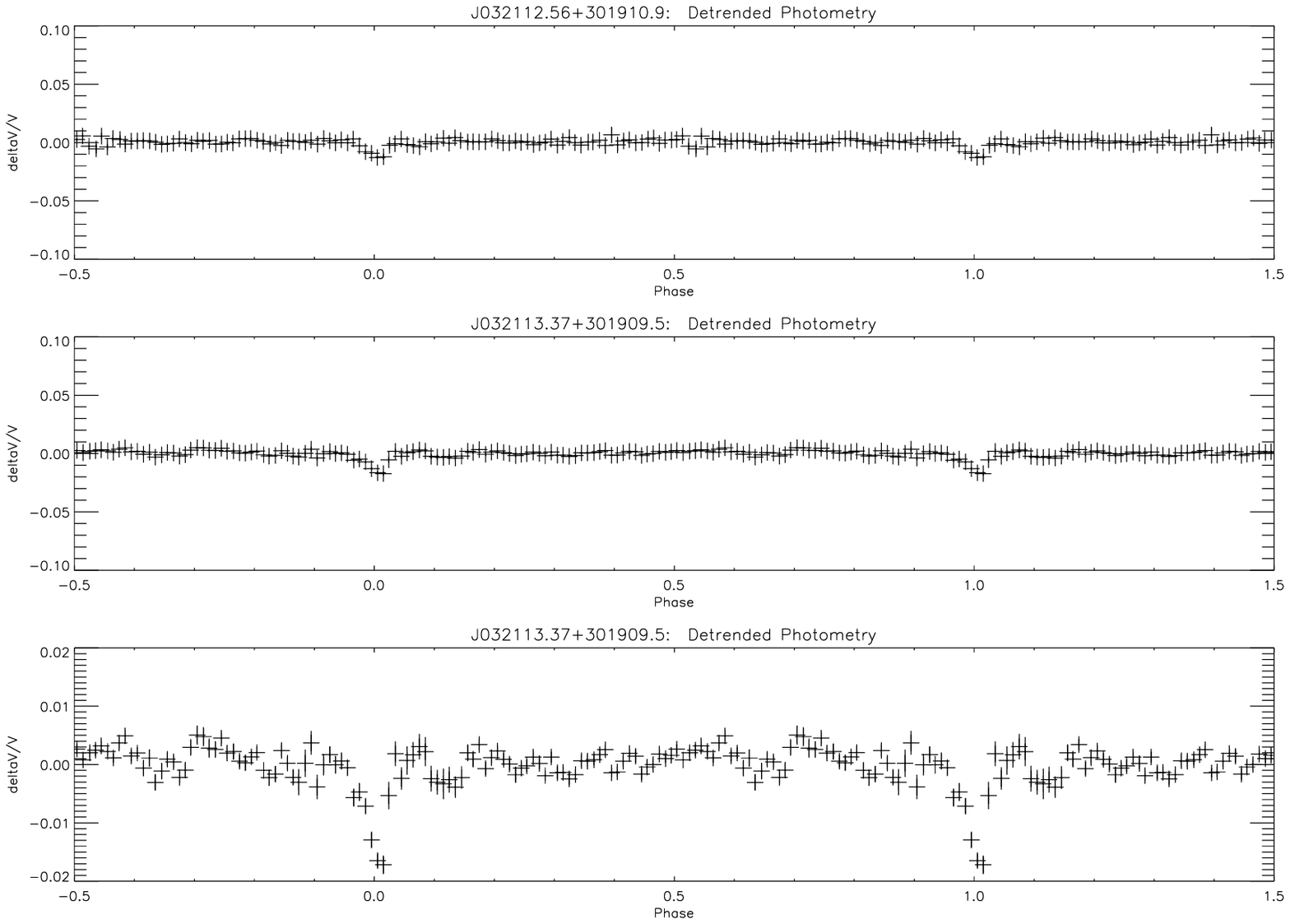,height=7cm}
}}
\caption{The visual double CCDM J031212+3019A \& B, recorded by the WASP pipeline as 1SWASP J032113.37+301909.5 \& J032112.56+301910.9 (Section 4.3). {\it Left:} Digitised Sky Survey image of the $5'\times5'$ region surrounding the two objects, with the immediate region surrounding the target (inset lower-right: the inner $30''\times30''$) and the full region binned to $15''\times15''$ pixels, approximately matching the SW-N pixel scale; Section 2.1 (inset upper-left). North is up, East to the left. The three SW-N apertures used for blending tests (2.5, 3.5 \& 4.5 pixels) are denoted by concentric circles. At $\sim 12''$ separation, the two components of the visual binary are so close that the resulting source extension under SW-N pixellation in the binned image falls entirely within the innermost aperture; thus the curve-of-growth blending index based on the three SW-N apertures misses the resulting blend. {\it Right:} Lightcurves of the two objects found by {\it SExtractor} during the pipeline reduction. At 2.27 days the recurrence interval is rather short for a fully detached stellar binary; however, little indication is found for ellipsoidal variations (bottom right).}
\label{visbinary}
\end{figure*}

{\bf J032113.37+301909.5 \& J032112.56+301910.9 - visual double:} Two separate objects are reported by the WASP pipeline with
similar lightcurves and periodograms, and positions $\sim12''$
apart. 
%Although only $\sim 2/3$ SW-N pix apart, the initial
%source-finding with {\it SExtractor} produced two separate peaks which
%were each reduced with the WASP pipeline. 
The closeness on-sky of the two lightcurves threw immediate suspicion
on either of these objects as planet-host candidates, which was
confirmed at the stage of catalogue examination. Four catalogues of
visual doubles provide matches with this object; the Couteau catalogue
of 2700 doubles (Couteau 1995), the Washington Visual Double Star
catalogue (Worley \& Douglas 1997), the CCDM (Dommanget et al. 2002)
and the Tycho Double Star Catalogue (Fabricius et al. 2002); this
object is the visual double CCDM J03212+3019A \& B. The primary is
listed as spectral type A5 (luminosity class not determined), and
objects A \& B have V-magnitudes 10.4 and 12.5 respectively. The
recurrence interval for the transit-like events is 2.26742(5) days,
which is rather short for a fully detached binary; however the
significance of any ellipsoidal variation is low (at S/N of
ellipsoidal variations $\la 1.7$; see also the lightcurve in Figure
\ref{visbinary}). One possible scenario is that the fainter object B
may be deeply eclipsed by a third, unseen object.  Although only $\sim
2/3$ SW-N pixels apart, the two components do produce an extended
object under a $15''$ pixel-scale (Figure \ref{visbinary}); initial
source-finding with {\it SExtractor} (Bertin \& Arnouts 1996)
localised this to two separate peaks which were each reduced with the
WASP pipeline. Although clearly a blend, this object did not fall
within the locus of blended objects based on comparison of flux within
the three SW-N apertures, because the object extension falls almost
entirely within the innermost aperture (Figure
\ref{visbinary}). Further examination of this object will be reported
elsewhere.

\section{Discussion}

Of a total of 141,895 targets extracted for the transit-search in the
fields considered here, 2688 were selected as potential photometric
transit candidates at the initial selection by $S_{red}$ and cadence
by the BLS algorithm (Collier Cameron et al. 2006). Of these, 44
passed the visual tests. The subsequent statistical tests removed all
but 20 of the candidates, of which 4 passed tests imposed by existing
catalogue photometry at higher spatial resolution; this last stage led
to the demotion of two otherwise Priority 1 objects. One object was
passed forward as a Priority 1 candidate for follow-up with other
facilities, with three more flagged as Priority 2 possibilities. As
the other WASP candidate-lists produced thus far (Christian et
al. 2006, Lister et al. 2007, Street et al. 2007) have resulted in
$\sim$3-4 times as many good candidates as the fields we report here,
it is worth examining the expected planet yield for the 03--06h RA
range.

% move this to right before the simulation plot

%\begin{figure}
% \psfig{file=visbinary.ps,width=6cm}
% \caption{{\it 2MASS} image of the visual binary consisting of SAO 56371 \&
%   CCDM J03212+3019B. Filter-glints, marked by ``X'', highlight the
%   visual separation of the two objects. SW-N + pipeline has detected
%   two objects, one of which is centered on the centre of the circle
%   in the diagram, the other of which coincides with the NE edge of
%   the binary.}
%\label{binary}
%\end{figure}

\begin{figure}
%\centerline{\hbox{
\psfig{file=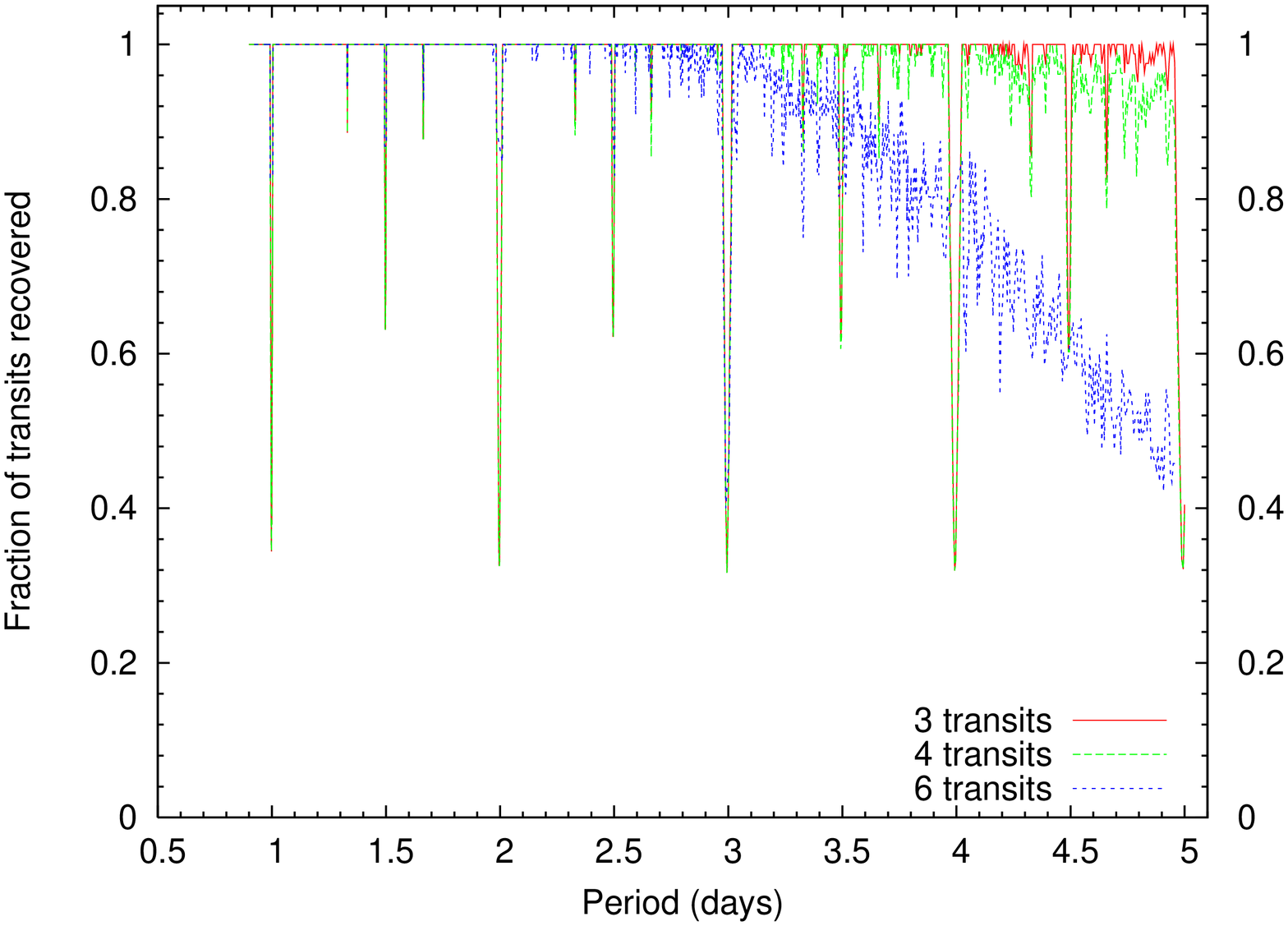,width=8cm,angle=-90}
\psfig{file=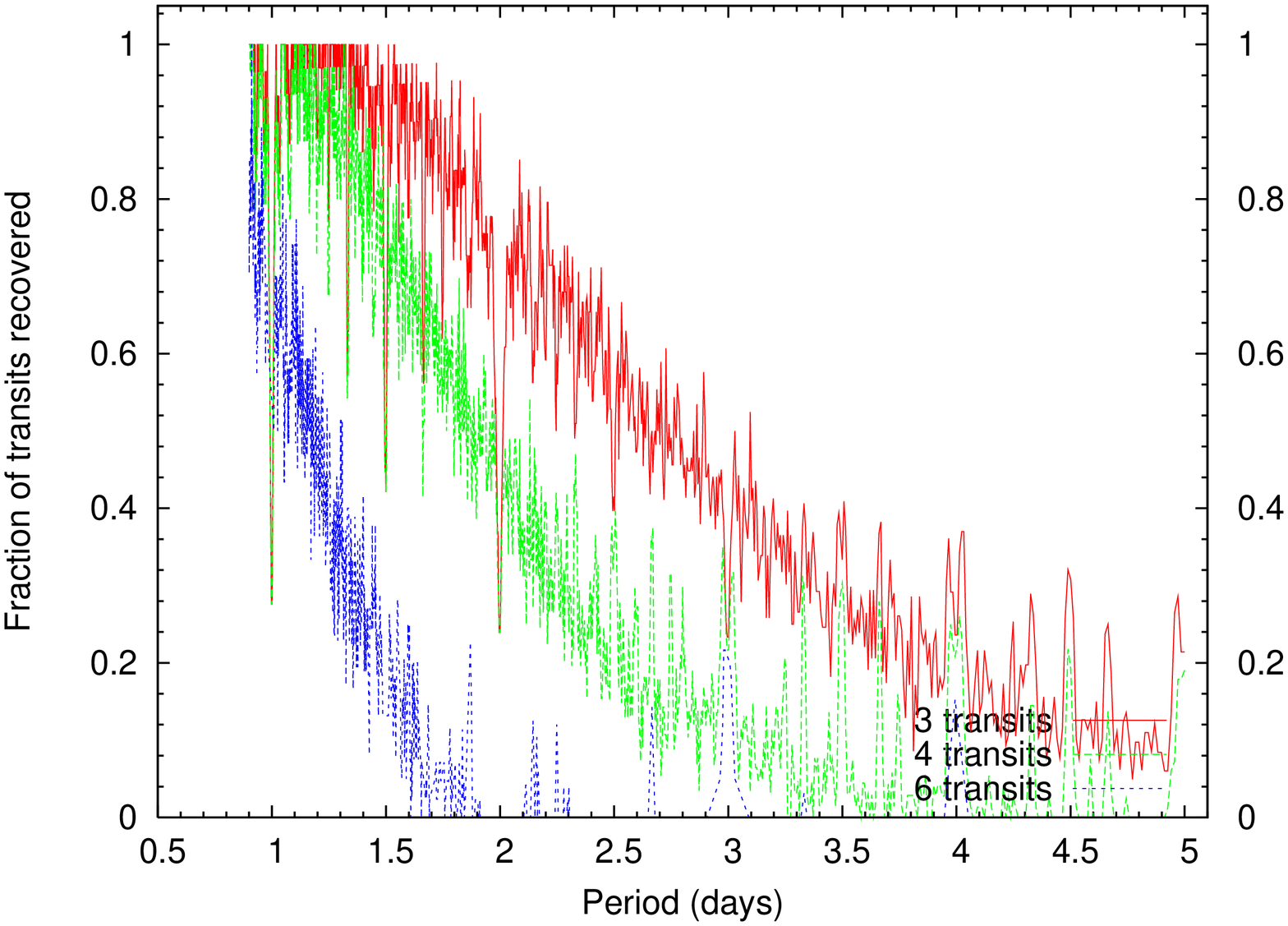,width=8cm,angle=-90}
\psfig{file=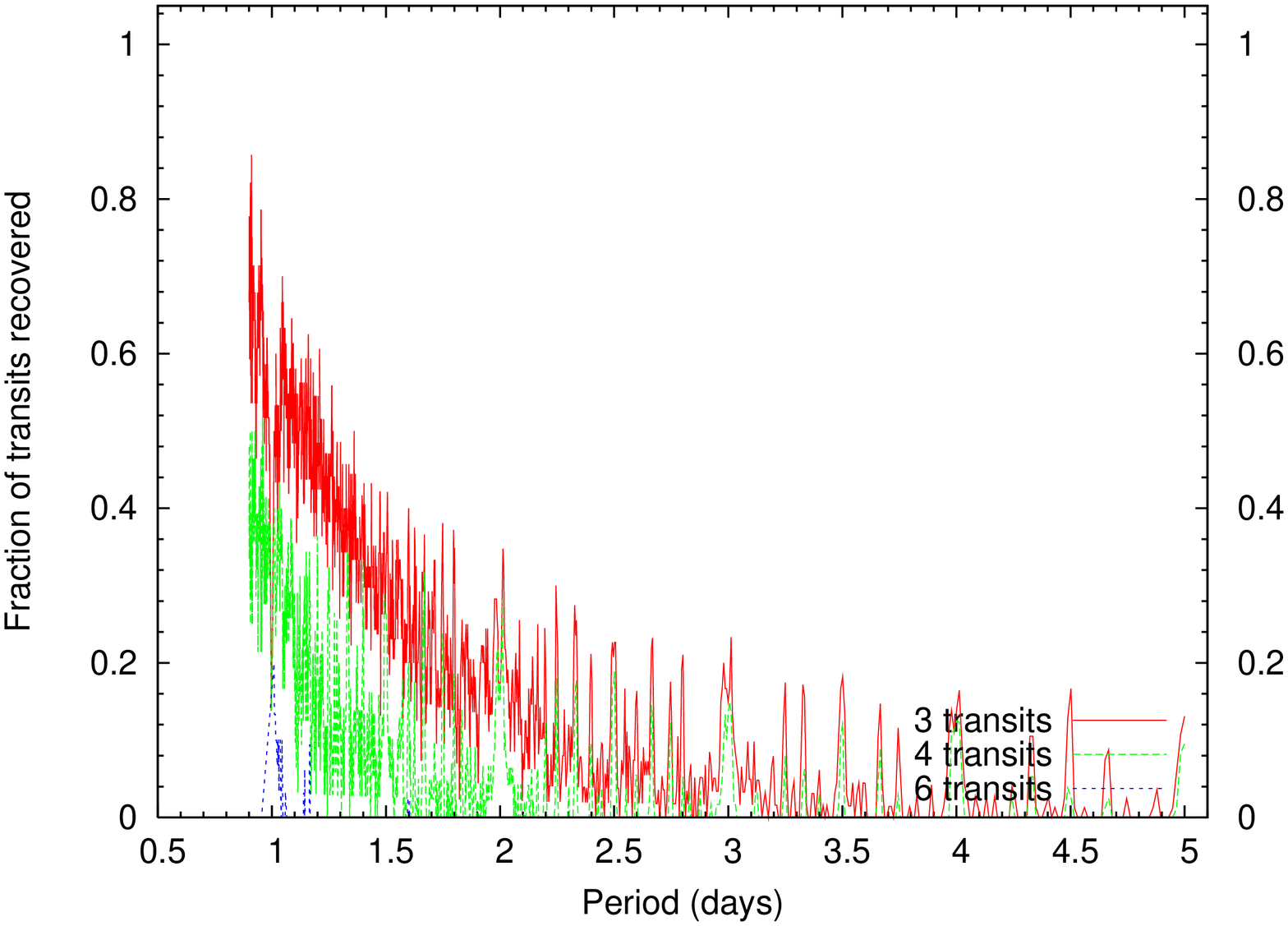,width=8cm,angle=-90}
%}}
\caption{Probability of transit detection as a function of planetary orbital period, for three (solid lines), four (dashed lines) and six (dotted lines) transits (see Collier Cameron et al. 2006). The coverage for these fields is as follows: {\bf Top:} 5441 frames over 129 nights (field 2045+1628); {\bf Middle:} 1402 frames over 64 nights (field SW0343+3126), {\bf Bottom:} 544 frames over 37 nights (SW0543+3126). There is a marked gradient in observability of transits with the number of nights observed; for 37 nights of data, the recovery fraction at 4 transits drops to 10 percent at all periods $\ga$ 1.5 days.}
\label{coverage}
\end{figure}

The relative dearth of transit candidates reported here is almost
certainly a result of the comparatively sparse sampling for this RA
range; the most intensively-observed field here consisted of 1885
frames over 60 nights, compared to e.g. 5541 frames over 127 nights
for a field on the other side of the sky (c.f. Street et
al. 2007). This has two key effects on our ability to detect
transits. The first effect was predicted before the survey began: even
with purely uncorrelated noise and an ideal instrument, the rotation
of the Earth imposes period-ranges in regions about integer-day
periods, within which the likelihood of dectecting transits is
reduced. As the Earth orbits the Sun and the sky precesses throughout
the year, the width of these intervals is reduced. These
low-observability windows are superimposed on a general decrease in
probability of transit observability with period due to fewer numbers
of long-period cycles falling within a typical observing season. To
clarify this point we present example estimates of the probability of
observing $N$ or more transits in a single SW-N observing season,
computed for each field as a byproduct of the transit search (Collier
Cameron et al. 2006), with the true sampling of each field as an input
(Figure \ref{coverage}).
% to simulate the impact on transit observability produced by
%the timing windows alone. Square-wave transit signals are randomly
%generated with uniformly random dates of first mid-eclipse and periods
%following a probability distribution matched to the observed
%hot-Jupiter period distribution from radial-velocity studies. Each
%artificial transit is then considered ``observed'' if both ingress and
%egress take place while the facility is active on-sky. 
%Figure \ref{coverage} shows the typical transit observability computed
%under the WASP-N observing strategy, using the actual
%magnitude-distribution of observed stars in each field and the true
%sampling for each field. 
As can be seen, below about 60 nights'
data-length, the recoverability of transits drops dramatically for all
but the shortest periods.

The second key effect of short observing timescales is the loss of
sensitivity in the presence of strong variability from correlated
noise, in which the noise power is not independent of the timescale of
variability. When planning ground-based transit searches, it was
largely assumed that improved reduction techniques would result in
uncorrelated noise (e.g. Horne 2003). In practice, despite the fact
that the magnitude of noise variation from several transit-surveys
(ours included) approach the Poisson floor for the entire magnitude
range over which we are sensitive to transits (here 8 $\la V \la 13$),
correlated noise continues to be a significant source of potential
false-positives, with significant power to variations with $\sim2.5$h
duration (similar to a genuine exoplanet transit). The only
ground-based broad-shallow transit-search team we are currently aware
of that claims uncorrelated noise is the XO group, which employs
drift-scanning to smooth out systematic trends instrumentally
(McCullough et al. 2006); thus it appears that correlated noise of
this nature may be a feature of the observational strategy we have
chosen.
% SHOULD ADD SOMETHING IN HERE TOO!
%{\it The obvious thing to do here is to take lightcurves of pure
%  noise-sources and do a straight LS periodogram of them. Binning this
%  up should make the frequency-dependence of the noise in the
%  lightcurves screamingly obvious. This is done routinely for
%  long-term X-ray datasets.}
The full signal to noise statistic $S_{red}$ of Pont et al. (2006)
provides a measure of the signal to noise of a transit detection in
the presence of correlated noise, and thus provides a useful
measurement to investigate ways to tame frequency-dependent correlated
noise. In particular we note the following two key results from the
considerations of Pont et al (2006): (i) that for ground-based transit
surveys the threshold to detect transits in the presence of correlated
noise is typically a factor $\sim3$ higher than in the presence of
uncorrelated noise alone, and that (ii) in the presence of correlated
noise, $S_{red}$ should scale roughly linearly with the total number
of nights of observation. The latter is particularly important, and
can be easily understood in the following way: consider a transit
signal with transit-duration $\delta_t$ and period $P_t$. The presence
of correlated noise with significant power at timescales
$\sim\delta_t$ will add spurious transit-like events at randomised
phases, reducing the coherence of the resulting transit lightcurve and
making the true periodicity more difficult to separate out from the
noise. Although comparison of the badness-of-fit allowing a model
consisting of both positive and negative transit-like events with that
from negative transits only (c.f. Burke et al. 2006) can to some
extent estimate the impact of correlated noise on the lightcurve
itself, it is clear that characterisation of the true transit period
in this case really requires as long an observing season as possible.

Indeed, simulations applied specifically to the WASP project (Smith et
al. 2006) suggest $S_{red}$ should increase roughly linearly with the
number of nights of observation; Smith et al (2006) suggest that
$\sim19 \pm 8$ genuine detections should be expected from the entire
WASP-N 2004 dataset; at $\la 60$ nights the recovery fraction drops to
roughly a quarter of that expected for datasets of length $\sim120$
nights. Our yield of four candidates is consistent with this scaling.

\section{Conclusions}

One Priority 1 exoplanet transit candidate has been uncovered from the
03--06h RA fields in the WASP-N 2004 dataset and three Priority 2
objects. This number is lower than that produced by other fields with
longer observation timebases. This is certainly due to the
comparatively sparse sampling, which bears out in a qualitative way
the results of recent work on correlated noise in ground-based
photometric surveys. When the 2006 SuperWASP datasets are fully
reduced, we expect to find many more candidates for follow-up work as
a result of the longer baseline this allows.

\section*{Acknowledgments}
%%\begin{acknowledgements} 
%\acknowledgements %  
The WASP consortium consists of representatives from the Queen's
University Belfast, University of Cambridge (Wide Field Astronomy Unit), 
Instituto de Astrofisica de Canarias,  Isaac Newton Group of Telescopes (La Palma),  University of
Keele, University of Leicester, Open University, and the University of St Andrews. 
The SuperWASP-North and South instruments were constructed and operated 
%The SuperWASP and WASP-S instruments were constructed and operated 
with funds made available from Consortium Universities   
%from Queen's University Belfast,  
and the Particle Physics and Astronomy Research Council. 
%and the Open University. 
SuperWASP I is located in the Spanish Roque de Los Muchachos Observatory on La Palma,  
Canary Islands which is operated by the Instituto de Astrofísica de Canarias (IAC). 
Several large astronomical catalogues were used via the {\it VizieR} service, operated at CDS in Strasbourg, France, primarily the Tycho-2 and USNO catalogues.  This publication makes use of data products from the Two Micron All Sky Survey, which is a joint project of the University of Massachussetts and the Infrared Processing and Analysis Center/California Institute of Technology, funded by the National Aeronautics and Space Administration and the National Science Foundation. The Digitized Sky Surveys were produced at the Space Telescope Science Institute under U.S. Government grant NAG W-2166. The images of these surveys are based on photographic data obtained using the Oschin Schmidt Telescope on Palomar Mountain and the UK Schmidt Telescope. The plates were processed into the present compressed digital form with the permission of these institutions. We have made use of the Extrasolar Planets Encyclopaedia, maintained on-line by Jean Schneider at http://exoplanet.eu/catalog.php.

We thank the anonymous referee for a number of insightful comments that greatly improved this manuscript. WIC thanks Peter McCullough and Kailash Sahu for fruitful discussion.

%%\end{acknowledgements}
%\vskip 1cm
%\noindent{\bf {\large References}}

%\include{references2.tex}
%REFERENCES 
%\begin{thebibliography}{} 
%\bibliographystyle{mn2e}
%\bibliography{iau_journals,master,ownrefs}
\bibliographystyle{mn2e}

\label{lastpage}
\end{document}